\documentclass[aps,pra,preprint,onecolumn,superscriptaddress]{revtex4-1}
\usepackage{amssymb,amsmath}
\usepackage{bm}
\usepackage{subfigure}
\usepackage{setspace}
\usepackage{booktabs}
\usepackage{mathrsfs}
\usepackage{graphicx}
\usepackage{color,soul}
\usepackage{xr}

\def\ben{\begin{equation}}
\def\een{\end{equation}}

\DeclareMathAlphabet\mathbfcal{OMS}{cmsy}{b}{n}
\newcommand{\mr}[1]{\mathrm{#1}}
\newcommand{\mc}[1]{\mathcal{#1}}

\newcommand{\mb}[1]{\mathbf{#1}}

\newcommand{\up}{_{\mr{p}}}

\def\up{\uparrow}
\def\dn{\downarrow}

\begin{document}

\title{Time-Dependent Coupled-Cluster Theory of\\ Multireference Systems}

\author{Mart\'in A. Mosquera}
\email{martinmosquera@montana.edu}
\affiliation{Department of Chemistry and Biochemistry, Montana State University, Bozeman, MT 59717,
USA}

\begin{abstract}  
In this work we present a coupled-cluster theory for the propagation of multireference electronic
systems initiating at general quantum mechanical states. Our formalism is based on the infinitesimal
analysis of modified cluster operators, from which we extract a set of additional operators and
their equations of motion. These cluster operators then can be used to compute electronic properties
of interest in a size extensive manner. The equations derived herein are also studied for the free
propagation of linear superpositions. In this regime, we derive asymmetric matrix elements for
observables, and a resymmetrization factor so they become consistent with their quantum mechanical
counterparts. Our formalism is applied to a four-electron/four-level model with a Hubbard-like
Hamiltonian, where we show that it reproduces the numerically exact results for observables,
populations, and coherences.
\end{abstract}

\maketitle
\section{Introduction}
The parameter-free computation of time-dependent (TD) quantum mechanical observables of electronic
systems that manifest strongly correlated effects are among the outstanding challenges in condensed
matter physics and quantum chemistry. This is due to their demand for accurate solutions to quantum
mechanical equations of a very high complexity. Progress in these directions could unlock needed
computational methods to understand the dynamics of discrete multi-spin systems, complex materials,
and related strongly correlated systems. These challenges combined have led to the emergence of
sophisticated approaches in the fields of quantum Monte Carlo
\cite{lee2022twenty,qian2025deep,mazzola2024quantum}, density-matrix renormalization group
\cite{lee2022twenty,mazzola2024quantum}, and coupled-cluster (CC)
\cite{qiu2017projected,guo2018communication} theories. For other strongly correlated systems such as
many-body quantum simulators \cite{browaeys2020many,gross2017quantum} and atom-based quantum
computers \cite{weiss2017quantum,henriet2020quantum}, advanced electronic structure theories are
also needed to quantify in a systematically improvable way the measurable properties of atomic
qubits.

In the fields of time-independent atomic and molecular physics, the multi-reference (MR)
self-consistent field \cite{hinze1973mc,dalgaard1978optimization,roos1980complete,
olsen2011casscf,olsen1988determinant,siegbahn1981complete} approach is critical to quantify and
understand optical, thermodynamic, and magnetic properties. Starting from an expansion of
determinantal wavefunctions, MR techniques generalize Hartree-Fock theory, and are able to account
for the strongly correlated character of these systems. However, they do not describe their full
electronic correlation, and the number of determinants can grow exponentially with system size,
which motivates advanced approaches mentioned above.  For description of full
electron-correlation effects, MR theories are complemented with methods that examine the missing
so-called weak correlation, or dynamical correlation. Notable examples of such complements are
perturbation theory (PT) \cite{andersson1990second,finley1998multi,lischka2018multireference},
configuration interaction (CI) \cite{shavitt1998history,craig1950configurational,
ross1952calculations,meyer1977configuration,werner1982self,sivalingam2016comparison,lechner2024code},
and coupled-cluster (CC) theory
\cite{coester1958bound,coester1960short,vcivzek1966correlation,vcivzek1969use,monkhorst1977calculation,
emrich1981extension,emrich1981extension2,mukherjee1979response,ghosh1984use,stanton1993equation,zhang2019coupled,
monkhorst1977calculation,schonhammer1978time,hoodbhoy1978time,hoodbhoy1979time,arponen1983variational}.
Roughly speaking, practical CI amounts to truncated diagonalization of Hamiltonians. CC, on the
other hand, while it truncates operators, it also extrapolates their effects to the complete
particle excitation limit.  In contrast to CI and CC methods, PT is more direct in the sense
that it provides analytical formulas for all observables, but this may come at the expense of
losing numerical stability to some degree. 

CC theories and their TD extensions
\cite{folkestad2023entanglement,sverdrup2023time,skeidsvoll2022simulating,
pathak2022time,pathak2021time,skeidsvoll2020time,pedersen2020interpretation,pedersen2019symplectic,
koulias2019relativistic,white2019time,white2018time,park2019equation,nascimento2017simulation,nascimento2016linear,
pigg2012time,walz2012application,kats2011second,sonk2011td,kvaal2012ab,sato2018communication,pathak2021time},
are highly appealing and commonly used in quantum chemistry (especially in non-Hermitian form), and recently in quantum computing
\cite{taube2009rethinking,shen2017quantum,harsha2018difference,lee2018generalized,
evangelista2019exact,romero2018strategies,xia2020qubit,ryabinkin2018qubit,anand2022quantum}, but are
also emerging in the study of crystals \cite{wang2021absorption,neufeld2022ground}.  They originate
in nuclear physics with the work of Coester \cite{coester1958bound,coester1960short} and the
extension to electronic systems by {\v{C}}{\'\i}{\v{z}}ek
\cite{vcivzek1966correlation,vcivzek1969use}, and have evolved into a robust set of computational
methodologies. CC techniques enjoy widespread use in treating dynamically correlated electrons,
systematically converge towards exact results \cite{pavsteka2017relativistic,reimann2022spin}, and
obey the principle of size extensivity \cite{hirata2011,hirata2014mutual}. To describe full
correlation (strong and dynamic), effective MR CC (MRCC) theories have been formulated
\cite{evangelista2018perspective,
jeziorski2010multireference,maitra2012unitary,hanrath2008multi,hanauer2011pilot,kohn2013state,
samanta2014excited,jagau2012linear,chattopadhyay2000development}. These theories differ from one
another in the way they treat the ground-state wavefunction, but the objective is the same, which is
transforming a ground state reference wavefunction into a more accurate one in general. This is
carried out by either splitting the operator that performs the transformation, or using a single
operator whose action on the multireference generates contracted wavefunctions; this latter approach
is known as internally contracted MRCC (ic-MRCC). In more detail, ic-MRCC theories rely on the
generation of intermediately normalized wavefunctions through a single exponential operator that
depends on electron-hole excitation operators, and equations that heavily feature commutators of
quantum mechanical operators. Excitation operators take advantage of orthogonality to the highest
extent possible. The presence of commutators leads to computing quantities that feature size
extensiveness, and where numerical results are systematically improvable.

In TD CC theories it is known how to compute the response of a system that starts at the ground
state, and in a size-extensive manner, but propagating general initial states is an open challenge.
This motivates this work, which formulates an internally contracted TD MRCC theory, referred to as
`extended TD-MRCC' (eTD-MRCC), to propagate MR systems that initiate at states such as entangled
states, linear superpositions, and pure excited states. eTD-MRCC is based on a theoretical approach
developed before by the author, named `second response theory' \cite{mosquera2022excited,
mosquera2023second}, which draws analogy from the formalism of infinitesimal generators that involve
the exponential operator. By considering initial states with infinitesimal contributions from
non-ground states, and using an action functional and its resulting equations of motion, we extract
an extended family of TD equations that are able to propagate general initial states. The present
work is based on non-Hermitian operators, and investigates the exact computation of TD quantum
mechanical observables. From the analysis of unperturbed non-stationary states and linear response
theory, we propose that the problem of non-symmetric transition matrix elements can be remedied
through the inclusion of resymmetrization factors that are related to CC wavefunction normalization.
The developed theoretical framework is studied in the context of a numerically exact model where
we show the possibility of numerically reproducing its TD observables, populations, and coherences.

\section{Theory}
We develop the formalism for non-relativistic (Coulombic) multi-electron systems. The TD MRCC theory
to propagate general initial states is formulated as follows: In Section \ref{motivation} we
motivate the concept that a small contribution from a non-ground state of interest can be added to
conventional CC operators that satisfy the conventional TD CC equations. This is equivalent to
adding extended CC operators (their representation is discussed in Section \ref{exci_ops}) that are
multiplied by an `infinitesimal' number. Through differential analysis, motion equations of these
extended CC operators are derived, including the equation to determine TD observables in our theory
with these operators, Section \ref{motion_eqs}. Then, we show the steps required to construct the
initial conditions of the extended operators, Sections \ref{td_obs} and \ref{unperturbed_TD_states},
and how static matrices, TD populations, and coherences are computed in our approach, Section
\ref{matrix_elem}.

\subsection{Definitions and Eigenstate Formalism}\label{motivation}
Several definitions are introduced: We denote $\langle\hat{O}\rangle$ as $\langle
0|\hat{O}|0\rangle$, where $\hat{O}$ is a quantum mechanical operator, and $|0\rangle$ is a given MR
state. The object $|0\rangle$ may derive from a method such as CASSCF (complete active space
self-consistent field), ground-state ic-MRCC, or a related approach to generate MR wavefunctions.
Atomic units are used throughout this work. For a ground-state cluster operator, $\hat{T}$, we
define the symbol $\bar{O}_T=\exp(-\hat{T})\hat{O}\exp(+\hat{T})$. Cluster operators considered in
this work are non-commutative. We denote $\partial_x$ as the partial derivative $\partial/\partial
x$. If $\hat{z}(\theta)$ is a cluster operator that depends on a continuous variable $\theta$, we
will use in this work the identities for derivatives of exponentials:
\ben
\begin{split}
\exp[-\hat{z}(\theta)](\partial_{\theta}\exp[+\hat{z}(\theta)]) &=
\hat{\pi}_z\partial_{\theta}\hat{z}~,\\
 (\partial_{\theta}\exp[-\hat{z}(\theta)])\exp[+\hat{z}(\theta)] &=
-\hat{\pi}_z\partial_{\theta}\hat{z}~,
\end{split}
\een
where $\hat{\pi}_z$ is an operator defined as:
\ben
\begin{split}
\hat{\pi}_z\cdot &= 1-\frac{1}{2}[\hat{z},\cdot]+\frac{1}{3!}\big[\hat{z},[\hat{z},\cdot]\big]
-\cdots\\
&=\sum_{n} \frac{(-1)^n}{(n+1)!}\mr{ad}_z^n\cdot~,
\end{split}
\een
where $\mr{ad}_z\cdot = [\hat{z},\cdot]$.
So $\hat{\pi}_z\partial_{\theta}\hat{z}=\partial_{\theta}\hat{z}-
(1/2)[\hat{z},\partial_{\theta}\hat{z}]+\cdots$. For a different cluster operator $\hat{u}$
depending on two continuous variables, $\alpha$ and $\beta$, we have that
\ben\label{liedev}
\partial_{\alpha}(e^{-\hat{u}}\partial_{\beta}e^{\hat{u}})
-\partial_{\beta}(e^{-\hat{u}}\partial_{\alpha}e^{\hat{u}})
=
[\hat{\pi}_u\partial_{\beta}\hat{u},\hat{\pi}_u\partial_{\alpha}\hat{u}]~.
\een
Second response theory \cite{mosquera2022excited,mosquera2023second} is motivated by the following
considerations: First, we introduce a wavefunction where the GS (ground state) is slightly perturbed
by our initial state of interest:
\ben
|\Psi_{\mr{R}}(g_{\mr{R}})\rangle = |\Psi_0\rangle+g_{\mr{R}}|\Psi(0)\rangle~,
\een
where $|\Psi_0\rangle$ is the GS of the system, $g_{\mr{R}}$ is a small (continuous) number, and
$|\Psi(0)\rangle$ is the initial state we wish to propagate. The present CC formalism is
non-Hermitian, this requires that we also consider the state $\langle \Psi_{\mr{L}}(g_{\mr{L}})| =
\langle\Psi_0|+g_{\mr{L}}\langle\Psi(0)|$, where $g_{\mr{L}}$ is a small number as well.

Let us assume for now that $|\Psi(0)\rangle=|\Psi_I\rangle$, where $|\Psi_I\rangle$ is a given ES
(excited state) eigenfunction of the $\hat{H}_0$ Hamiltonian operator. 
Then it is evident that
\ben\label{I_state}
\frac{\partial}{\partial g_{\mr{R}}} |\Psi_{\mr{R}}\rangle = |\Psi_I\rangle~.
\een
The analogue of this relation can be pursued in MRCC theory. Let us introduce
\ben
|\Phi_{\mr{m}}(g_{\mr{R}})\rangle = \exp[\hat{T}+g_{\mr{R}}(\hat{X}^I-\mr{i}\phi_I)]|0\rangle~,
\een
where $\hat{X}^I$ is a cluster operator, and $\phi_I$ is a phase, 
which can be taken as a purely imaginary number. This phase factor is
required to satisfy biorthogonality with left-handed GS CC wavefunctions, see Section \ref{matrix_elem}.  
Because $|\tilde{\Phi}_{\mr{m}}\rangle$ is non-linear with respect to $g_{\mr{R}}$, and motivated by Eq.
(\ref{I_state}), we demand that in the limit where this variable tends to zero,
\ben
\lim_{g_{\mr{R}}\rightarrow 0}\frac{\partial}{\partial g_{\mr{R}}} |\tilde{\Phi}_{\mr{m}}(g_{\mr{R}})\rangle
=\exp(\hat{T})\hat{\mc{R}}_I|0\rangle
\een
gives a CC wavefunction related to the state $|\Psi_I\rangle$, 
where 
\ben
\hat{\mc{R}}^I = \hat{\pi}_T\hat{X}^I-\mr{i}\phi_I~.
\een
The free Hamiltonian of the system is denoted as $\hat{H}_0$. If exact,
$e^{\hat{T}}\hat{\mc{R}}^I|0\rangle$ must satisfy
$\hat{H}_0\exp(\hat{T})\hat{\mc{R}}^I|0\rangle=E_I\exp(\hat{T})\hat{\mc{R}}^I|0\rangle$.  This is
equivalent to writing the problem as
$\bar{H}_T^0\hat{\mc{R}}^I|0\rangle=E_I\hat{\mc{R}}^I|0\rangle$, where
$\bar{H}^0_T=e^{-\hat{T}}\hat{H}_0e^{\hat{T}}$. Similarly, the exact GS CC wavefunction solves the
eigenvalue problem $\bar{H}_T^0|0\rangle=E_0|0\rangle$. Combining the ES and GS eigenvalue
equations, the ES problem can be expressed as:
\ben\label{r_eig}
[\bar{H}_T^0,\hat{\pi}_T\hat{X}^I]|0\rangle = \Omega_I\hat{\pi}_T\hat{X}^I|0\rangle~,
\een
where $\Omega_I=E_I-E_0$ is the excitation energy of the system (from GS to ES labeled ``I'').  The
above equation is exact if the problem is solved to all electron excitation orders. In the single-reference
case, it becomes the conventional EOM-CC (equation-of-motion CC) problem because
$\hat{\pi}_T\hat{X}^I=\hat{X}^I$ (a consequence of their commutative excitation operators). In the
MR domain, the operator $\hat{\pi}_T$ arises from the lack of commutativity between cluster operators
($\hat{T}$ and $\hat{X}^I$ do not commute, for example).  In computational implementations, the
above equation can be expressed directly as a matrix algebra problem, which would be based on the
projection $\langle\hat{\tau}^{\dagger}_N [\bar{H}_T^0,\hat{\pi}_T\hat{X}^I]\rangle = \Omega_I
\langle \hat{\tau}^{\dagger}_N\hat{\pi}_T\hat{X}^I\rangle$, where $\hat{\tau}_N$ is an excitation
operator, defined in Section \ref{exci_ops}.

For the left-handed problem, we consider the wavefunction $\langle\Psi_{\mr{L}}(g_{\mr{L}})|=
\langle\Psi_0|+g_{\mr{L}}\langle\Psi(0)|$. Suppose $\langle\Psi(0)|$ is $\langle\Psi_I |$, and define the CC
analogue: $\langle \tilde{\Upsilon}_{\mr{m}}(g_{\mr{L}})| =\langle
0|(\hat{\lambda}_0+g_{\mr{L}}\hat{\Lambda}^I)\exp(-\hat{T})$ (a phase is not required here
because biorthogonality is satisfied with this form), where $\langle
0|\hat{\lambda}_0\exp(-\hat{T})$ is the left CC representation of the GS wavefunction, 
and $\hat{\Lambda}^I$ a conjugate cluster operator. 
Differentiating with respect to $g_{\mr{L}}$
gives the state $\langle \Psi_I|$ and its CC analogue $\langle 0|\hat{\Lambda}^I\exp(-\hat{T})$, the
formal expression $\langle 0|\hat{\Lambda}^IE_I=\langle 0|\hat{\Lambda}^I\bar{H}_T^0$ leads to 
\ben
\langle \hat{\Lambda}^I[\bar{H}_T^0,\hat{\pi}_T\hat{\tau}_N]\rangle=\Omega_I\langle
\hat{\Lambda}^I\hat{\pi}_T\hat{\tau}_N\rangle~.
\een 
This equation represents a generalized eigenvalue problem, where the biorthogonality condition reads
$\langle\hat{\Lambda}^I\hat{\pi}_T\hat{X}^J\rangle=\delta_{IJ}$.

Let us consider a general initial state of the form $|\Psi(0)\rangle=c_0|\Psi_0\rangle+\sum_{J>0}c_J|\Psi_J\rangle$, 
where $\{c_I\}$ denote the standard linear superposition coefficients,
and define the standard quantum mechanical TD observable, whose TD CC computation is an objective of this
work:
\ben
\langle \hat{B}\rangle_{\Psi}(t)=\langle \Psi(t)|\hat{B}|\Psi(t)\rangle~,
\een
where $|\Psi(t)\rangle=\hat{U}(t)|\Psi(0)\rangle$, $\hat{U}(t)$ is the unitary evolution
operator of the system, $t$ represents time, and $\hat{B}$ is the operator corresponding to that observable of interest. 
Formally, $\hat{U}(t)=\mc{T}\exp[-\mr{i}\int_0^t\hat{H}(s)\mr{d}s]$, where $\mc{T}$ is the
time-ordering superoperator, and $\hat{H}(t)$ the TD Hamiltonian operator of the system.
Additionally, we define:
\ben
\mc{B}(t;g_{\mr{L}},g_{\mr{R}}) =
\frac{\langle\Psi_{\mr{L}}|\hat{U}^{\dagger}(t)\hat{B}\hat{U}(t)|\Psi_{\mr{R}}\rangle}
{\langle \Psi_{\mr{L}}|\Psi_{\mr{R}}\rangle}~.
\een
In previous work \cite{mosquera2023second,mosquera2022excited}, we proved the following identity that
connects the standard observable with its CC equivalent in our formalism:
\ben\label{qm_b_obs}
\langle B(t)\rangle_{\Psi} = \lim_{g_{\mr{L}},g_{\mr{R}}\rightarrow 0}
\Big\{
  \frac{\partial^2}{\partial g_{\mr{L}} \partial g_{\mr{R}}} \mc{B}(t)
  +c_0\frac{\partial}{\partial g_{\mr{L}}} \mc{B}(t)
  +c_0^*\frac{\partial}{\partial g_{\mr{R}}} \mc{B}(t)
  +\langle \Psi_0|\hat{U}^{\dagger}(t)\mc{B}\hat{U}(t)|\Psi_0\rangle
  \Big\}~.
\een
The limit operation is redundant in this last expression, but it is convenient for the CC theory
to be developed in the rest of this work, as it relates the average value $\langle
\hat{B}\rangle_{\Psi}(t)$ to the sum of first and second order derivatives of $\mc{B}(t)$ with
respect to the variables $g_{\mr{L}}$ and $g_{\mr{L}}$, and the average value of $\hat{B}$ when the
system is propagated from the GS.

\subsection{Excitation Operators}\label{exci_ops}
In this subsection we describe the generation of excitation operators that allow us to focus on a TD
MRCC approach where the time-dependency is accounted for by cluster operators exclusively (other
choices to generate excitation operators are possible for this eTD-MRCC formalism). Let
$|\mr{MR}\rangle$ be a MR wavefunction, with respect to this state ($|\mr{MR}\rangle$),
conventional excitation operators are denoted as $\tilde{\tau}_{\mu}$ (these are products of
electron annihilation/creation operators), an intermediate set of excitation operators will be
referred to as $\{\hat{\tau}_N'\}$, and our re-defined transformed operators are denoted as
$\{\hat{\tau}_N\}$. All possible conventional particle excitations are considered in this procedure. 

We introduce the overlap matrix
$S_{\mu\nu}=\langle\mr{MR}| \tilde{\tau}^{\dagger}_{\mu}\tilde{\tau}_{\nu}|\mr{MR}\rangle$. This
matrix includes the multireference state, so $\tilde{\tau}_{0}=1$ for convenience.  We use solution
to the singular-value decomposition (SVD) problem for positive semi-definite matrix, so
$\mb{U}^{\dagger}\mb{S}\mb{U} = \bm{I}$, where $\mb{U}=\mb{V}\bm{\Sigma}^{-1/2}$, $\mb{\Sigma}$ is
the diagonal matrix of singular values, and $\mb{V}\bm{\Sigma}\mb{V}^{\dagger}=\mb{S}$. Only
columns of $\mb{U}$ with singular value greater than a threshold $\eta$ $(\Sigma_{NN} > \eta$) are
included to generate excitation operators. Using the non-singular contractions one obtains, with
respect to $\mb{S}$, a set of excitation operators, $\hat{\tau}'_N=\sum_{\mu}U_{\mu
N}\tilde{\tau}_{\mu}$ ($N\ge 0$), such that $\langle
(\hat{\tau}_N')^{\dagger}\hat{\tau}_N'\rangle=\delta_{MN}$. The reference state, $|\mr{MR}\rangle$,
is not a member of this set. However, it can be written as $|\mr{MR}\rangle = \sum_N d_N|N\rangle$,
where $|N\rangle = \hat{\tau}_N'|\mr{MR}\rangle$ and $d_N=\langle N|\mr{MR}\rangle =
\sum_{\mu}U^{\dagger}_{N \mu}S_{\mu 0}$. Now we define $b_{MN} = d_M$ if $N=0$, and $b_{MN} =
\delta_{MN}$ otherwise ($N>0$). Column vectors in $\mb{U}$ are organized in decreasing order with
respect to their singular values, except $\hat{\tau}'_0|\mr{MR}\rangle$, which is taken as the state
with the highest overlap with $|\mr{MR}\rangle$, and the first column vector in the $\mb{U}$ matrix.
A QR factorization of $\mb{b}$ allows us to write this matrix as $\mb{b}=\mb{q}\mb{r}$, and produces
a new set of transformed operators
\ben
\hat{\tau}_N=\sum_{\mu} Q_{\mu N}\tilde{\tau}_{\mu}~,
\een
where $Q_{\mu N} = \sum_M U_{\mu M} q_{MN}$,
so that $\hat{\tau}_0 = 1$ and $|0\rangle = \hat{\tau}_0|\mr{MR}\rangle = |\mr{MR}\rangle$; 
leading to an orthonormal basis $\{\hat{\tau}_N |0\rangle\}_{N\ge 0}$, where $|\mr{MR}\rangle$ is
indeed included.

Using the above definitions, the GS cluster operators are expressed as $\hat{T}=\sum_{N>0}
t_N\hat{\tau}_N$, and $\hat{\Lambda}_0=\sum_{N>0}\hat{\tau}^{\dagger}_N\Lambda_{0,N}$.  Now we apply
the variational principle \cite{helgaker1988analytical,helgaker1989numerically} to the ground-state
MRCC problem: $\mc{E}(\mb{t},\bm{\Lambda}_0)=\langle
(1+\hat{\Lambda}_0)\exp(-\hat{T})\hat{H}_0\exp(+\hat{T})\rangle$, which is the conventional
GS CC problem to obtain the cluster amplitudes. First, $\langle
\hat{\tau}^{\dagger}_N\exp(-\hat{T})\hat{H}_0\exp(+\hat{T})\rangle = 0$, and $\langle
(1+\hat{\Lambda}_0)[\bar{H}^0_T,\hat{\pi}_T\hat{\tau}_N]\rangle = 0$. Because two different contracted
operators $\hat{\tau}_N$ and $\hat{\tau}_M$ do not commute, $\partial_{t_N}\bar{H}_0$
does not have a compact expression, but it can be truncated (as an approximation) for direct
computation, or it can be determined numerically by other means. 

\subsection{Motion Equations}\label{motion_eqs}
The time dependency of the states $\hat{U}(t)|\Psi_{\mr{R}}(g_{\mr{R}})\rangle$ and $\langle
\Psi_{\mr{L}}(g_{\mr{L}})|\hat{U}^{\dagger}(t)$ offers all the information needed to study the
propagation of a general initial state $\Psi(0)$, Refs. \cite{mosquera2022excited,
mosquera2023second}. This can be pursued within TD CC theory as well. We denote our CC-analogue
objects to $\hat{U}(t)|\Psi_{\mr{R}}\rangle$ and $\langle \Psi_{\mr{L}}|\hat{U}^{\dagger}(t)$ as
$|\Phi_{\mr{m}}(t; g_{\mr{R}})\rangle$ and $\langle\Upsilon_{\mr{m}}(t; g_{\mr{L}}, g_{\mr{R}})|$,
correspondingly; the ``$\mr{m}$'' subscript stands for ``modified''. In terms of
operators, we express them as $|\Phi_{\mr{m}}(t)\rangle =
\exp[\hat{x}_{\mr{m}}(t)-\mr{i}\phi_{\mr{m}}(t)]|0\rangle$, and the left TD CC wavefunction
$\langle\Upsilon_{\mr{m}}(t)|=\langle
0|\hat{\lambda}_{\mr{m}}(t)\exp[-\hat{x}_{\mr{m}}(t)+\mr{i}\phi_{\mr{m}}(t)]$
\cite{koch1990coupled,hansen2019time}. To study the time-evolution of these objects, we introduce
the following TD cluster operators, which are central to our theory: $\hat{x}_{\mr{r}}(t)$,
$\hat{\lambda}_{\mr{l}}(t)$, $\hat{\lambda}_{\mr{r}}(t)$, $\hat{\lambda}_{\mr{lr}}(t)$, and
$\phi_{\mr{r}}(t)$. These operators are similar to the well-known `generators' used in continuous
group theory, but they have not been used in TD CC theory to the best of our knowledge. 
In addition, we also introduce standard TD CC operators and phase associated to
propagation from the GS exclusively, these are $\hat{x}(t)$, $\hat{\lambda}(t)$, and $\phi(t)$,
where $\hat{x}(0)=\hat{T}$, $\hat{\lambda}(0)=1+\hat{\Lambda}_0$, and $\phi(0)=0$. The initial
conditions of the modified cluster operators are:
\ben
\begin{split}
\hat{x}_{\mr{m}}(0;g_{\mr{R}})&= \hat{x}(0) + g_{\mr{R}}\hat{x}_{\mr{r}}(0)~,\\
\hat{\lambda}_{\mr{m}}(0;g_{\mr{L}},g_{\mr{R}})&=\hat{\lambda}(0) + g_{\mr{L}}\hat{\lambda}_{\mr{l}}(0) +
g_{\mr{R}}\hat{\lambda}_{\mr{r}}(0) +  g_{\mr{L}}g_{\mr{R}}\hat{\lambda}_{\mr{lr}}(0)~.
\end{split}
\een
For the phase we have that $\phi_{\mr{m}}(t=0) = \phi(0)+g_{\mr{R}}\phi_{\mr{r}}(0)$.
We interpret the new operators as follows:
The operator $\hat{x}_{\mr{r}}(t)$ and the phase $\phi_{\mr{r}}(t)$ are related to the ES component
of $|\Psi(t)\rangle$, $\hat{\lambda}_{\mr{l}}$ is related to $\langle \Psi(t)|$. The operators
$\hat{\lambda}_{\mr{r}}(t)$ and $\hat{\lambda}_{\mr{lr}}(t)$ are, in part, corrective terms that arise
from the non-Hermitian nature of the formalism, but (for mathematical convenience) they also contain
information about propagation from the GS. This interpretation is used to construct the initial states
of these operators, and to express their connection to quantum mechanical observables, section
\ref{td_obs}.

Now, let us consider the action functional (time-interval implicit) \cite{hansen2019time,mosquera2022excited}:
\ben
\begin{split}
\mc{F}[\mb{x}_{\mr{m}}, \bm{\lambda}_{\mr{m}}, \phi_{\mr{m}}]&=\int\mr{d}t~\langle
\Upsilon_{\mr{m}}(t)|\hat{H}(t)-\mr{i}\vec{\partial}_t|\Phi_{\mr{m}}(t)\rangle\\
&=\int\mr{d}t~\big\langle\hat{\lambda}_{\mr{m}}(t)\big\{\exp[-\hat{x}_{\mr{m}}(t)]\hat{H}(t)\exp[+\hat{x}_{\mr{m}}(t)]-\mr{i}
\hat{\pi}_{x_{\mr{m}}}(t)\partial_t\hat{x}_{\mr{m}}(t)-\partial_t\phi_{\mr{m}}(t)\big\}\big\rangle~.
\end{split}
\een
where $\hat{x}_{\mr{m}}(t)=\sum_{N>0}x_{\mr{m},N}(t)\hat{\tau}_N$ and
$\hat{\lambda}_{\mr{m}}(t)=\sum_{N}\hat{\tau}_N^{\dagger}\lambda_{\mr{m},N}(t)$; $\mb{x}_m$ and 
$\bm{\lambda}_{\mr{m}}$ represent the vectors $\{x_{\mr{m},N}(t)\}$ and $\{\lambda_{\mr{m},N}(t)\}$ as a
whole. 
The variational principle and its associated Euler-Lagrange relations give the equation:
\ben\label{dt_xm}
\mr{i}\langle \hat{\tau}_N^{\dagger}\hat{\pi}_{x_{\mr{m}}}(t)\partial_t\hat{x}_{\mr{m}}\rangle=\langle
\hat{\tau}^{\dagger}_N\hat{H}_{x_{\mr{m}}}(t)\rangle~,
\een
where $\hat{H}_{x_{\mr{m}}}(t)=\exp[-\hat{x}_{\mr{m}}(t)]\hat{H}(t)\exp[+\hat{x}_{\mr{m}}(t)]$. 

In an analogous way, we obtain the motion equation of the conjugate cluster operator:
\ben\label{dt_lamb_m}
-\mr{i}\langle \partial_t\hat{\lambda}_{\mr{m}}\hat{\pi}_{x_{\mr{m}}}(t)\hat{\tau}_N\rangle=
\langle\hat{\lambda}_{\mr{m}}(t)[\hat{H}_{x_{\mr{m}}}(t),\hat{\pi}_{x_{\mr{m}}}(t)\hat{\tau}_N]\rangle
+\mr{i}\langle\hat{\lambda}_{\mr{m}}(t)[\hat{\pi}_{x_{\mr{m}}}(t)\hat{\tau}_N,\hat{\pi}_{x_{\mr{m}}}(t)\partial_t\hat{x}_{\mr{m}}]\rangle~.
\een
The last term follows from Eq. (\ref{liedev}) being used here, and avoids computation of derivative of
$\hat{\pi}_{x_{\mr{m}}}$ with respect to time. 
In the left hand side of the equation shown above, $\partial_t$ only applies to
$\hat{\lambda}_{\mr{m}}$. Regarding the modified TD phase, setting $\mc{F}=0$ leads to the relation
$\partial_t\phi_{\mr{m}}=\langle \hat{\lambda}_{\mr{m}}(t)\hat{H}_{x_{\mr{m}}}(t)\rangle$.  This
equation is derived by noticing that $\langle \Upsilon_{\mr{m}}(t)|\Phi_{\mr{m}}(t)\rangle = 1$,
which implies that
$\langle\hat{\lambda}_{\mr{m}}(t)\hat{\pi}_x(t)\partial_t\hat{x}_{\mr{m}}\rangle=0$, this property
and Eq. (\ref{dt_xm}) mean that the phase can also be expressed as $\partial_t\phi_{\mr{m}}=\langle
\hat{H}_{x_{\mr{m}}}(t)\rangle$. 
The TD equations for propagation from GS are obtained after making
$g_{\mr{R}}=0$ and $g_{\mr{L}}=0$, and removing the subscript `m' from Eqs. (\ref{dt_xm}) and
(\ref{dt_lamb_m}). Computing $\hat{x}(t)$ and $\hat{\lambda}(t)$ in this way allows for calculation of
any observable as $\langle \hat{\lambda}(t)\exp[-\hat{x}(t)]\hat{B}\exp[\hat{x}(t)]\rangle$
(provided the initial state is the GS).

Similarly as the $g_{\mr{R}}$-derivative of $\hat{U}(t)|\Psi_{\mr{R}}\rangle$ gives the time-evolved
state $\hat{U}(t)|\Psi(0)\rangle$, we can differentiate too the modified cluster operators to obtain
information about that propagation. As formulated before \cite{mosquera2022excited,
mosquera2023second} for single reference propagation (but the same principle applies to this case),
we have that:
\ben
\begin{split}
\hat{x}_{\mr{r}}(t) &=\lim_{g_{\mr{R}}\rightarrow 0} \frac{\partial}{\partial
g_{\mr{R}}} \hat{x}_{\mr{m}}(t),~
\hat{\lambda}_{\mr{l}}(t) = \lim_{g_{\mr{R}},g_{\mr{L}}\rightarrow 0}\frac{\partial
  }{\partial g_{\mr{L}}}\hat{\lambda}_{\mr{m}}(t),~
\hat{\lambda}_{\mr{r}}(t) =  \lim_{g_{\mr{R}},g_{\mr{L}}\rightarrow 0}
  \frac{\partial}{\partial g_{\mr{R}}}\hat{\lambda}_{\mr{m}}(t),\\
\hat{\lambda}_{\mr{lr}}(t) &=  \lim_{g_{\mr{R}},g_{\mr{L}}\rightarrow 0}
  \frac{\partial^2}{\partial g_{\mr{L}}\partial g_{\mr{R}}} \hat{\lambda}_{\mr{m}}(t),~
\phi_{\mr{r}}(t) =  \lim_{g_{\mr{R}},g_{\mr{L}}\rightarrow 0} \frac{\partial
}{\partial g_{\mr{R}}}\phi_{\mr{m}}(t)~.
\end{split}
\een
Differentiating the modified TD equations with respect to
$g_{\mr{R}}$ or/and $g_{\mr{L}}$, and taking the limit $g_{\mr{L}}, g_{\mr{R}}\rightarrow 0$, we
obtain the following:
\ben\label{xr_eq}
\mr{i}\langle\hat{\tau}_N^{\dagger} \hat{\pi}_x(t)\partial_t\hat{x}_{\mr{r}}\rangle = \langle
\hat{\tau}^{\dagger}_N[\hat{H}_x(t),\hat{\pi}_x(t)\hat{x}_{\mr{r}}(t)]\rangle
- \mr{i}\langle\hat{\tau}_N^{\dagger} \hat{\pi}_{\mr{r}}'(t) \partial_t\hat{x}\rangle~,
\een
where $\hat{H}_x(t)=\exp[-\hat{x}(t)]\hat{H}(t)\exp[+\hat{x}(t)]$, and 
\ben
\hat{\pi}_{\mr{r}}'(t)=\lim_{g_{\mr{R}}\rightarrow 0}\frac{\partial }{\partial
g_{\mr{R}}}\hat{\pi}_{x_{\mr{m}}}(t)~.
\een
The object $\hat{\pi}_{\mr{r}}'(t)$ can be computed with a recursion formula, Appendix \ref{pir_section}.
For the operator $\hat{\lambda}_{\mr{l}}$ we find that:
\ben\label{lambl_eq}
-\mr{i}\langle\partial_t\hat{\lambda}_{\mr{l}}\hat{\pi}_x(t)\hat{\tau}_N\rangle=\langle\hat{\lambda}_{\mr{l}}(t)[\hat{H}_x(t),\hat{\pi}_x(t)\hat{\tau}_N]\rangle
+\mr{i}\langle\hat{\lambda}_{\mr{l}}(t)[\hat{\pi}_{x}(t)\hat{\tau}_N,\hat{\pi}_{x}(t)\partial_t\hat{x}]\rangle~.
\een
The mixed term $\hat{\lambda}_{\mr{lr}}$ is the solution to the relation below:
\ben\label{lamb_lr}
\begin{split}
-\mr{i}\langle\partial_t\hat{\lambda}_{\mr{lr}}\hat{\pi}_x(t)\hat{\tau}_N\rangle
=&\langle\hat{\lambda}_{\mr{lr}}(t)[\hat{H}_x(t),\hat{\pi}_x(t)\hat{\tau}_N]\rangle+
\langle\hat{\lambda}_{\mr{l}}(t)\big[[\hat{H}_x(t),\hat{\pi}_x(t)\hat{x}_{\mr{r}}(t)],\hat{\pi}_x(t)\hat{\tau}_N\big]\rangle\\
&+\langle\hat{\lambda}_{\mr{l}}(t)[\hat{H}_x(t),
\hat{\pi}_{\mr{r}}'(t)\hat{\tau}_N]\rangle
+\Gamma_{\mr{lr},N}(t)~.
\end{split}
\een
This last equation is required by the formalism as it ensures consistency
for the calculation of observable properties in general. The $\Gamma_{\mr{lr},N}(t)$ term reads:
\ben
\begin{split}
-\mr{i}&\Gamma_{\mr{lr},N}(t) = 
\langle \hat{\lambda}_{\mr{lr}}(t)[\hat{\pi}_x(t)\hat{\tau}_N,\hat{\pi}_x(t)\partial_t\hat{x}]\rangle+
\langle \hat{\lambda}_{\mr{l}}(t)[\hat{\pi}_{\mr{r}}'(t)\hat{\tau}_N,\hat{\pi}_x(t)\partial_t\hat{x}]\rangle\\
&+\langle \hat{\lambda}_{\mr{l}}(t)[\hat{\pi}_x(t)\hat{\tau}_N,\hat{\pi}_{\mr{r}}'(t)\partial_t\hat{x}]\rangle+
\langle \hat{\lambda}_{\mr{l}}(t)[\hat{\pi}_x(t)\hat{\tau}_N,\hat{\pi}_x(t)\partial_t\hat{x}_{\mr{r}}]\rangle
+\langle \partial_t\hat{\lambda}_{\mr{l}}\hat{\pi}_{\mr{r}}'(t)\hat{\tau}_N\rangle~.
\end{split}
\een
The motion equation for $\hat{\lambda}_{\mr{r}}$ is similar to that of
$\hat{\lambda}_{\mr{lr}}$, and is shown in Appendix \ref{appdx_lambr}. Regarding the phase
$\phi_{\mr{r}}(t)$, from differentiation of
$\partial_t\phi_{\mr{m}}(t)=\langle\hat{H}_{x_{\mr{m}}}(t)\rangle$ in the limit
$g_{\mr{R}}\rightarrow 0$, we find the equation: $\partial_t\phi_{\mr{r}}(t)=\langle
[\hat{H}_{x}(t), \hat{\pi}_x(t)\hat{x}_{\mr{r}}(t)]\rangle$.

\subsection{Time-Dependent Observables and Initial States}\label{td_obs}

The CC expectation value that is analogous to $\langle
\Psi_{\mr{L}}|\hat{U}^{\dagger}(t)\hat{B}\hat{U}(t)|\Psi_{\mr{R}}\rangle/\langle\Psi_{\mr{L}}|\Psi_{\mr{R}}\rangle$ 
reads:
\ben
\tilde{B}(t) = \frac{\langle
\hat{\lambda}_{\mr{m}}(t)e^{-\hat{x}_\mr{m}(t)}\hat{B}e^{+\hat{x}_{\mr{m}}(t)}\rangle}{\mc{N}_{\mr{LR}}}~,
\een
where $\mc{N}_{\mr{LR}}=\langle \hat{\lambda}_{\mr{m}}(0)\rangle$; this norm is constant with
respect to time.  Although we are studying non-commutative cluster operators, the principal mathematical steps
are analogous to our previous work, Ref. \cite{mosquera2023second}. Applied to the present MR
study, it can be shown that a quantum mechanical observable is given by:
\ben\label{BPsi}
\langle \hat{B}\rangle_{\Psi}(t) = \lim_{g_{\mr{L}},g_{\mr{R}}\rightarrow 0}
\Big\{
  \frac{\partial^2}{\partial g_{\mr{L}} \partial g_{\mr{R}}} \tilde{B}(t)
  +C_0\frac{\partial}{\partial g_{\mr{L}}} \tilde{B}(t)
  +D_0\frac{\partial}{\partial g_{\mr{R}}} \tilde{B}(t)
  +\langle \hat{\lambda}_{\mr{m}}(t)\bar{B}_{x_{\mr{m}}}(t)\rangle
  \Big\}~.
\een
The terms $C_0$ and $D_0$ are linear superposition coefficients for
the GS, defined in Section \ref{matrix_elem}.
Equation (\ref{BPsi}) closely resembles its parent quantum mechanical version, [Eq. (\ref{qm_b_obs})], and it leads to 
the general formula:
\ben\label{gen_form}
\langle \hat{B}\rangle_{\Psi}(t) = 
\langle \hat{\lambda}_{\mr{l}}(t)[\hat{B}_{x}(t),\hat{\pi}_x(t)\hat{x}_{\mr{r}}(t)]\rangle
+\langle\hat{\lambda}_{\mr{lr}}(t)\hat{B}_{x}(t)\rangle~,
\een
where $\hat{B}_x(t)=\exp[-\hat{x}(t)]\hat{B}\exp[+\hat{x}(t)]$.
To compute a TD observable, one thus needs to determine the cluster operators $\hat{x}(t)$,
$\hat{x}_{\mr{r}}(t)$, $\hat{\lambda}_{\mr{l}}(t)$, and $\hat{\lambda}_{\mr{lr}}(t)$. 
The operator $\hat{\lambda}_{\mr{r}}(t)$ is subsumed by $\hat{\lambda}_{\mr{lr}}(t)$
\cite{mosquera2023second}.

For convenience, we decompose the ``lambda'' cluster operators as shown below \cite{mosquera2023second}:
\ben\label{lambs_t}
\begin{split}
\hat{\lambda}_{\mr{l}}(t) &= \hat{\lambda}_{\mr{l}}^{\mr{E}}(t)+D_0\hat{\lambda}(t)~,\\
\hat{\lambda}_{\mr{r}}(t) &= \hat{\lambda}_{\mr{r}}^{\mr{E}}(t)+C_0\hat{\lambda}(t)~,\\
\hat{\lambda}_{\mr{lr}}(t) &= \hat{\lambda}_{\mr{lr}}^{\mr{E}}(t)+C_0\hat{\lambda}_{\mr{l}}^{\mr{E}}(t)
+\hat{\lambda}(t)+D_0\hat{\lambda}_{\mr{r}}^{\mr{E}}(t)~,
\end{split}
\een
where $\hat{\lambda}_{\mr{l}}^{\mr{E}}(t)$, $\hat{\lambda}_{\mr{r}}^{\mr{E}}(t)$,
and $\hat{\lambda}_{\mr{lr}}^{\mr{E}}(t)$ obey the same equations as 
$\hat{\lambda}_{\mr{l}}(t)$, $\hat{\lambda}_{\mr{r}}(t)$,
and $\hat{\lambda}_{\mr{lr}}(t)$, respectively, but they only contain information related to excited
states and their interaction. It is important to note that
$\hat{\lambda}_{\mr{l}}(t)$, $\hat{\lambda}_{\mr{r}}(t)$, and $\hat{\lambda}_{\mr{lr}}(t)$, carry
terms that multiply the unity operator (which come from the fact that 
$\hat{\lambda}(0)=1+\hat{\Lambda}_0$).
The initial conditions for the cluster operators $\hat{x}_{\mr{r}}$,
$\hat{\lambda}_{\mr{l}}^{\mr{E}}$, $\hat{\lambda}_{\mr{r}}^{\mr{E}}$,
and $\hat{\lambda}_{\mr{lr}}^{\mr{E}}$, are: 
\ben\label{l_init}
\begin{split}
\hat{x}_{\mr{r}}(0) &= \sum_{I>0} C_I\hat{X}^I~,\\
\hat{\lambda}_{\mr{l}}^{\mr{E}}(0)&=\sum_{I>0} D_I\hat{\Lambda}^I~,\\
\hat{\lambda}_{\mr{r}}^{\mr{E}}(0)&=\sum_{J>0} D_{\mr{r},J}\hat{\Lambda}^J~,\\
\hat{\lambda}_{\mr{lr}}^{\mr{E}}(0)&=\sum_{J>0} D_{\mr{lr},J}\hat{\Lambda}^J~,\\
\end{split}
\een
where all the sums are over excited states exclusively. Together with $\hat{x}(0)=\hat{T}$ and
$\hat{\lambda}(0)=1+\hat{\Lambda}_0$, Eq. (\ref{l_init}) gives the initial conditions for 
$\hat{\lambda}_{\mr{l}}$, $\hat{\lambda}_{\mr{r}}$,
and $\hat{\lambda}_{\mr{lr}}$ after setting $t=0$ in Eq. (\ref{lambs_t}).
For the derivative phase,
$\phi_{\mr{r}}(0)=\sum_{I>0}C_I\phi_I$, where $\mr{i}\phi_I =
\langle\hat{\Lambda}_0\hat{\pi}_T\hat{X}^I\rangle$, Section \ref{matrix_elem}.  These expressions,
Eq. (\ref{l_init}), are helpful because they eliminate the need to solve the equation for
$\hat{\lambda}_{\mr{r}}(t)$, so only its initial condition is required. They also reproduce the
initial value of an observable of interest. In the next subsections, \ref{unperturbed_TD_states} and
\ref{matrix_elem}, we describe computation of the (complex-valued) coefficients, $\{C_I\}$,
$\{D_I\}$, $\{D_{\mr{r},J}\}$, and $\{D_{\mr{lr},J}\}$.

\subsection{Unperturbed Time-Dependent States}\label{unperturbed_TD_states} 

Here we derive expressions for the solution of the motion equations in absence of external fields,
so $\hat{H}(t)=\hat{H}_0$. They provide information necessary to perform driven propagations with
our eTD-MRCC approach. 
We denote the Jacobian matrix as $(\mb{A})_{NM} =
\langle\hat{\tau}^{\dagger}_N[\bar{H}_T^0,\hat{\tau}_M]\rangle$. In operator form, the eigenstates of this matrix are
taken as $\hat{X}^I=\sum_{N>0} X^I_N\hat{\tau}_N$, and $\hat{\Lambda}^I=\sum_{N>0}
\Lambda^I_N\hat{\tau}_N^{\dagger}$ ($I>0$).
It can be noted that linear combinations of TD CC operators of the form
$\hat{x}_{\mr{r}}(t)=\sum_{I>0} C_I\hat{X}^I\exp(-\mr{i}\Omega_I t)$ and
$\hat{\lambda}_{\mr{l}}^{\mr{E}}(t)=\sum_{I>0} D_I \hat{\Lambda}^I \exp(+\mr{i}\Omega_I t)$ obey the equations
$\mr{i}\bm{\pi}_T\partial_t\mb{x}_{\mr{r}}=\mb{A}\mb{x}_{\mr{r}}$ and
$-\mr{i}(\partial_t\bm{\lambda}_{\mr{l}}^{\mr{E}})\bm{\pi}_T=\bm{\lambda}_{\mr{l}}^{\mr{E}}\mb{A}$, if 
\ben
\begin{split}
\mb{A}\mb{X}^I&=\Omega_I\bm{\pi}_T\mb{X}^I~,\\
\mb{A}^{\dagger}\bm{\Lambda}^I&=\Omega_I\bm{\pi}_T^{\dagger}\bm{\Lambda}^I~, 
\end{split}
\een
where $(\bm{\pi}_T)_{NM}=\langle\hat{\tau}_N^{\dagger}\hat{\pi}_T\hat{\tau}_M\rangle$, 
$[\mb{x}_{\mr{r}}(t)]_N=\langle \hat{\tau}^{\dagger}_N\hat{x}_{\mr{r}}(t)\rangle$,
$[\bm{\lambda}_{\mr{l}}^{\mr{E}}(t)]_N^{\mr{T}}=\langle\hat{\lambda}_{\mr{l}}^{\mr{E}}(t)\hat{\tau}_N\rangle$,
and $\mb{X}^I,~\bm{\Lambda}^I$ refer to the vectors $\{X^I_N\}~,\{\Lambda^I_N$\}, correspondingly.
So we identify the above equation as the generalized eigenvalue problem to solve. Furthermore, for a standard
initial wavefunction superposition:
\ben 
|\Psi(0)\rangle=\sum_I c_I|\Psi_I\rangle~,
\een 
where $\{|\Psi_I\rangle\}$ are the eigenstates of the Hamiltonian $\hat{H}_0$, we have that $C_I$ is
directly related to $c_I$, and $D_I$ to $c_I^*$, as discussed in Section \ref{matrix_elem}.

In absence of perturbation, $\hat{x}(t)=\hat{T}$, and $\hat{\lambda}(t)=\hat{\lambda}_0$, where
$\hat{\lambda}_0=1+\hat{\Lambda}_0$. Now we expand the cluster operator
as $\hat{\lambda}_{\mr{r}}^{\mr{E}}(t)=\sum_{J>0} \tilde{D}_{\mr{r},J}(t)\hat{\Lambda}^J$, where
$\{\tilde{D}_{\mr{r},J}(t)\}$ is a set of TD coefficients. 
The unperturbed equations are solved under the restriction that
the operator $\hat{\lambda}_{\mr{r}}^{\mr{E}}$ cannot introduce new frequencies different from those already
associated to the vectors $\hat{x}_{\mr{r}}$ and $\hat{\lambda}_{\mr{l}}^{\mr{E}}$. This demands that the
initial state must be chosen such that it cancels the non-physical frequencies. 

Using the expressions for the unperturbed $\hat{x}_{\mr{r}}$ and $\hat{\lambda}_{\mr{l}}^{\mr{E}}$, we will
show that:
\ben\label{eq_for_D}
\tilde{D}_{\mr{r},J}(t)=-\sum_{I>0} C_I\frac{F^{IJ}}{\Omega_I+\Omega_J}\exp(-\mr{i}\Omega_I t)~,
\een
where
\ben
F^{IJ} = \langle
\hat{\lambda}_0\big[[\bar{H}_T^0,\hat{\pi}_T\hat{X}^I],\hat{\pi}_T\hat{X}^J\big]\rangle
+\Omega_I\langle\hat{\lambda}_0[\hat{\pi}_T\hat{X}^J,\hat{\pi}_T \hat{X}^I]\rangle
+\langle\hat{\lambda}_0[\bar{H}_T^0,\nabla\hat{\pi}_T\cdot\mb{X}^I\hat{X}^J]\rangle~,
\een
and
\ben\label{nabpi_t}
\nabla \hat{\pi}_T\cdot \mb{X}^I = \sum_{N>0}\frac{\partial \hat{\pi}_u}{\partial
u_N}\Bigg|_{\hat{u}=\hat{T}}X_N^I~.
\een
This term is equivalent to $\exp(\mr{i}\Omega_I t)\hat{\pi}_{\mr{r}}'$ when
$\hat{x}_{\mr{r}}(t)=\hat{X}^I \exp(-\mr{i}\Omega_I t)$ and $\hat{x}(t)=\hat{T}$. 

Let us consider first the case where the system is described by the linear superposition of the
GS and one given ES, where that excited state is labeled ``I''. It is clear that:
\ben
|\Psi(t)\rangle = e^{-\mr{i}E_0 t}\big(c_0|\Psi_0\rangle+c_Ie^{-\mr{i}\Omega_I
t}|\Psi_I\rangle\big)~.
\een
Suppose that $c_I$ is small in relation to $c_0$, so $|c_I|^2$ can be neglected. Thus, we obtain:
\ben\label{un_obs}
\langle B\rangle_{\Psi}(t) = |c_0|^2\langle\Psi_0|\hat{B}|\Psi_0\rangle
+ c_0^*c_Ie^{-\mr{i}\Omega_It}\langle\Psi_0|\hat{B}|\Psi_I\rangle
+c_I^*c_0e^{+\mr{i}\Omega_It}\langle\Psi_I|\hat{B}|\Psi_0\rangle+o(|c_I|^2)~.
\een
To represent this state, we use the cluster operators
$\hat{x}_{\mr{r}}(t)=C_Ie^{-\mr{i}\Omega_It}\hat{X}^I$, 
$\hat{\lambda}_{\mr{l}}^{\mr{E}}(t)=\hat{\Lambda}^ID_Ie^{+\mr{i}\Omega_It}$, and the solution to
Eq. (\ref{r_eq}), which now reads (Appendix \ref{appdx_lambr}):
\ben
\begin{split}
-\mr{i}\langle \partial_t\hat{\lambda}_{\mr{r}}^{\mr{E}}\hat{\pi}_T\hat{\tau}_N\rangle =& 
\langle \hat{\lambda}_{\mr{r}}^{\mr{E}}(t)[\bar{H}_T^0,\hat{\pi}_T\hat{\tau}_N]\rangle
+ \langle
\hat{\lambda}_0\big[[\bar{H}_T^0,\hat{\pi}_T\hat{x}_{\mr{r}}(t)],\hat{\pi}_T\hat{\tau}_N\big]\rangle\\
+& \langle\hat{\lambda}_0[\bar{H}_T^0,\nabla\hat{\pi}_T\cdot\mb{X}^I\hat{\tau}_N]\rangle
+\mr{i}\langle\hat{\lambda}_0[\hat{\pi}_T\hat{\tau}_N,\hat{\pi}_T\partial_t\hat{x}_{\mr{r}}(t)]\rangle~.
\end{split}
\een
Multiplying the above equation by $X^I_N$, summing over all states, and solving the resulting
equation give:
\ben\label{dr_J}
\tilde{D}_{\mr{r},J}(t) = \tilde{D}_{\mr{r},0,J}e^{-\mr{i}\Omega_Jt}-
C_I\frac{F^{IJ}}{\Omega_I+\Omega_J}(e^{-\mr{i}\Omega_I t}-e^{-\mr{i}\Omega_J t})~.
\een
So for the observable we find that [Eq. (\ref{gen_form})]:
\ben
\begin{split}
\langle\hat{B}\rangle_{\Psi}(t)=
C_0&D_0\langle\hat{\lambda}_0\bar{B}_T\rangle
+D_0\Big[C_I\langle \hat{\lambda}_0[\bar{B}_T,\hat{\pi}_T\hat{X}^I]\rangle e ^{-\mr{i}\Omega_It}
-\sum_{J>0}\Big(\tilde{D}_{\mr{r},0,J}e^{-\mr{i}\Omega_Jt}\\
&-\frac{C_I F^{IJ}}{\Omega_I+\Omega_J}(e^{-\mr{i}\Omega_I t}-e^{-\mr{i}\Omega_J
t})\Big)\langle\hat{\Lambda}^J\bar{B}_T\rangle\Big]
+D_IC_0e^{+\mr{i}\Omega_It}\langle\hat{\Lambda}^I\bar{B}_T\rangle
+o(|c_I|^2)~.
\end{split}
\een
Comparing this equation with Eq. (\ref{un_obs}), we demand that the cluster operator
cannot introduce phasors different from $\exp(-\mr{i}\Omega_It)$, therefore
\ben
\tilde{D}_{\mr{r},0,J}=C_I\frac{F^{IJ}}{\Omega_I+\Omega_J}~.
\een
Equation (\ref{eq_for_D}) derives from generalizing this argument to a super superposition of multiple states.

When the system is described exclusively by a superposition of a given number of excited states,
the procedure to solve the unperturbed (free propagation) problem is very similar to that just
discussed above.  The cluster operators would take the form
$\hat{x}_{\mr{r}}(t)=\sum_{I>0}C_Ie^{-\mr{i}\Omega_It}\hat{X}^I$,
$\hat{\lambda}_{\mr{l}}(t)=\sum_{I>0}\hat{\Lambda}^ID_Ie^{+\mr{i}\Omega_It}$
(here $\hat{\lambda}_{\mr{l}}^{\mr{E}}=\hat{\lambda}_{\mr{l}}$, since $c_0=0$). The wavefunction
would also read
\ben
\Psi(t)=e^{-\mr{i}E_0t}\sum_Ic_Ie^{-\mr{i}\Omega_It}\Psi_I~,
\een
where $c_0=C_0=D_0=0$. 
We introduce $\hat{\lambda}_{\mr{lr}}^{\mr{E}}(t)=\sum_{J>0}
\tilde{D}_{\mr{lr},J}(t)\hat{\Lambda}^J$ (with $\{\tilde{D}_{\mr{lr},J}(t)\}$ being another set of
TD coefficients).
To solve for $\tilde{D}_{\mr{lr},J}$,
the unperturbed TD equation [Eq. (\ref{lamb_lr})] of interest becomes:
\ben
\begin{split}
-\mr{i}\langle \partial_t\hat{\lambda}_{\mr{lr}}^{\mr{E}}&\hat{\pi}_T\hat{\tau}_N\rangle =
\langle \hat{\lambda}_{\mr{lr}}^{\mr{E}}(t)[\bar{H}_T^0,\hat{\pi}_T\hat{\tau}_N]\rangle
+ \langle
\hat{\lambda}_{\mr{l}}^{\mr{E}}(t)\big[[\bar{H}_T^0,\hat{\pi}_T\hat{x}_{\mr{r}}(t)],\hat{\pi}_T\hat{\tau}_N\big]\rangle\\
+& \langle\hat{\lambda}_{\mr{l}}^{\mr{E}}(t)[\bar{H}_T^0,\nabla\hat{\pi}_T\cdot \mb{X}^I \hat{\tau}_N]\rangle +
\mr{i}\langle\hat{\lambda}_{\mr{l}}^{\mr{E}}(t)[\hat{\pi}_T\hat{\tau}_N,\hat{\pi}_T\partial_t\hat{x}_{\mr{r}}(t)]\rangle
+\mr{i}\langle \partial_t\hat{\lambda}_{\mr{l}}^{\mr{E}}\hat{\pi}_{\mr{r}}'\hat{\tau}_N\rangle~.
\end{split}
\een
The solution to this equation is analogous to the previous analysis, where one
expands the operators, multiplies both sides by $X^I_N$, and sums over all $N$. Once more, because
$\hat{\lambda}^{\mr{E}}_{\mr{lr}}(t)$ cannot give rise to extra phasors different from those
associated to the initial state, we obtain Eq. (\ref{dlr}). The mentioned steps lead to:
\ben\label{dlr}
\tilde{D}_{\mr{lr},J}(t)=
  \sum_{I,K > 0}D_KC_I\frac{\mathcal{G}_{KIJ}}{\Omega_K-\Omega_J-\Omega_I}\exp[-\mr{i}(\Omega_I-\Omega_K)t]~,
\een
where
\ben\label{gkij}
\begin{split}
\mc{G}_{KIJ}=
\langle
\hat{\Lambda}_K&\big[[\bar{H}_T^0,\hat{\pi}_T\hat{X}^I],\hat{\pi}_T\hat{X}^J\big]\rangle
+\Omega_I\langle\hat{\Lambda}^K[\hat{\pi}_T\hat{X}^J,\hat{\pi}_T\hat{X}^I]\rangle\\
&+\langle\hat{\Lambda}_K[\bar{H}_T^0,\nabla\hat{\pi}_T\cdot\mb{X}^I\hat{X}^J]\rangle
-\Omega_K\langle\hat{\Lambda}_K\nabla\hat{\pi}_T\cdot\mb{X}^I\hat{X}^J\rangle~.
\end{split}
\een
The amplitudes $\{D_{\mr{r},J}\}$ and $\{D_{\mr{lr},J}\}$ needed in Section \ref{td_obs} to
determine $\hat{\lambda}_{\mr{r}}(0)$ and $\hat{\lambda}_{\mr{lr}}(0)$ follow from Eqs. (\ref{eq_for_D})
and (\ref{dlr}) after setting $t=0$: 
\ben
\begin{split}
D_{\mr{r},J}&=\tilde{D}_{\mr{r},J}(0)~,\\
D_{\mr{lr},J}&=\tilde{D}_{\mr{lr},J}(0)~.
\end{split}
\een  
These are also required to compute matrix elements of quantum
mechanical properties, Section \ref{matrix_elem}.

\subsection{Matrix Elements and Their Exact Limit}\label{matrix_elem}
In the previous subsection we showed the identification of mathematical elements involved in quantum
mechanical property calculation, but it is ambiguous with regards to $C_I$ and $D_I$ coefficients.
To resolve this, we can examine the standard TD linear response (LR) of the system (with initial
state being the GS), following similar steps as in previous LR theories \cite{koch1990coupled}; Ref.
\cite{mosquera2022excited} discusses this in the single reference scenario.  In the LR regime, one
then finds that matrix elements can be extracted from the following product:
\ben
\begin{split}
f_{0I}(\hat{A})h_{I0}(\hat{B}) &= \Big(\langle \hat{\lambda}_0
[\bar{A}_T,\hat{\pi}_T\hat{X}^I]\rangle-\sum_{J>0} \frac{F^{IJ}}{\Omega_I+\Omega_I}
\langle \hat{\Lambda}^J\bar{A}_T\rangle\Big)\big(\langle\hat{\Lambda}^I\bar{B}_T\rangle\big)\\
&=\langle \Psi_0|\hat{A}|\Psi_I\rangle\langle \Psi_I|\hat{B}|\Psi_0\rangle~.
\end{split}
\een
LR theory suggests that this product is an exact representation, but it does not pinpoint the individual
matrix elements (a factor is missing). However, one can note the ratio:
\ben
\frac{f_{0I}(\hat{A})h_{I0}(\hat{B})}{f_{0I}(\hat{A})h_{I0}(\hat{C})} = \frac{\langle \Psi_I|\hat{B}|\Psi_0\rangle}
{\langle \Psi_I|\hat{C}|\Psi_0\rangle}=
\frac{\langle\hat{\Lambda}^I\bar{B}_T\rangle}{\langle\hat{\Lambda}^I\bar{C}_T\rangle}~,
\een
where a third operator, $\hat{C}$ is introduced. A similar result applies for the element associated
for $f_{0I}$. 

From the above observation, we hypothesize that the exact quantum mechanical elements are proportional to the CC
counterparts. To find expressions for these proportionality terms, we now assume that the CC
problem were solved to all particle excitation orders. In that limit, we obtain that
$f_{0I}(\hat{B})=\langle\hat{\lambda}_0\bar{B}_T(\hat{\pi}_T\hat{X}^I-\mr{i}\phi_I)\rangle $ and $h_{I0}(\hat{B})=
\langle \hat{\Lambda}_I\bar{B}_T\rangle$ (this is proven next). The phase $\phi_I$ derives from the
bi-orthogonaliry constraint: $\langle \hat{\lambda}_0\hat{\mc{R}}^I\rangle=0$, which implies that
\ben
\mr{i}\phi_I = \langle\hat{\Lambda}_0\hat{\pi}_T\hat{X}^I\rangle~.
\een 
States such as $e^{\hat{T}}\hat{\mc{R}}^I|0\rangle$ and $\langle
0|\hat{\Lambda}^Ie^{-\hat{T}}$ are not normalized from a standard quantum mechanical perspective. To
maintain consistency with standard QM results, we propose that the coefficients $\{C_I\}$ and
$\{D_I\}$ must reflect this normalization, so 
\ben\label{CI_DI}
C_I = \frac{c_I}{\mc{N}_{\mr{r},I}},~D_I = \frac{c_I^*}{\mc{N}_{\mr{l,I}}}~,
\een
\sloppy where the normalization factors read $\mc{N}_{\mr{r},0} = \lVert \exp(\hat{T})
\rVert$, $\mc{N}_{\mr{l},0} = \lVert[ \hat{\lambda}_0 \exp(-\hat{T})]^{\dagger} \rVert$,
$\mc{N}_{\mr{l},I} = \lVert [\hat{\Lambda}^I\exp(-\hat{T})]^{\dagger} \rVert$, and $\mc{N}_{\mr{r},I} = \lVert
\exp(\hat{T})(\hat{\pi}_T\hat{X}^I-\mr{i}\phi_I)\rVert$, for all $I$, where $\lVert\hat{O}\rVert^2
=\langle\hat{O}^{\dagger}\hat{O}\rangle$. A disadvantage of these relations is that they require
Hermitian conjugation of operators, but they offer a path towards an exact theory for TD
observables. 
Therefore, in our approach $\{C_I\}$ and $\{D_I\}$ are calculated with Eq. (\ref{CI_DI}), 
then used to compute the coefficients $\{D_{\mr{r},J}\}$ and $\{D_{\mr{lr},J}\}$, and
the initial conditions of the extended cluster operators, Eq. (\ref{l_init}). 

A conventional eigenstate $|\Psi_I\rangle$ is then treated as the CC state
$\mc{N}_{\mr{r},I}^{-1}\exp(+\hat{T})\hat{\mc{R}}_I|0\rangle$, and $\langle \Psi_I|$ as $\langle
0|\hat{\Lambda}^I\exp(-\hat{T})\mc{N}_{\mr{l},I}^{-1}$; this applies to the GS wavefunctions
$|\Psi_0\rangle$ and $\langle\Psi_0|$ if $\hat{\mc{R}}^0$ is taken as $1$ and $\hat{\Lambda}^0$ as
$\hat{\lambda}_0$, respectively. The product of reciprocal normalization factors is the unity, or
$\mc{N}_{\mr{r},I}\mc{N}_{\mr{l},I}=1$ (this also holds if $I=0$). One can choose the eigenvectors
to satisfy $\langle \hat{\Lambda}^I\hat{\pi}_T\hat{X}^I\rangle = 1$ ($I>0$), so when including the
CC-state normalization ($\langle\Psi_I|\Psi_I\rangle$):
$\langle\hat{\Lambda}^I\hat{\pi}_T\hat{X}^I\rangle/(\mc{N}_{\mr{l},I}\mc{N}_{\mr{r},I}) = 1$. These
relationships explain previous simulations \cite{mosquera2022excited,mosquera2023second} where we
have noticed that diagonal matrix elements of static observables are numerically exact, while the
off-diagonal terms are not.  Additionally, it supports observations in which our TD results for
general initial states are significantly close to the numerically exact reference data, but not the
same. In other words, the hidden normalization factors approximately cancel each other for
off-diagonal matrix elements. 

If the EOM-MRCC eigenvalue problem has been solved to all particle excitation orders, then
$\bar{H}^0_T\hat{\mc{R}}^I|0\rangle = E_I\hat{\mc{R}}^I|0\rangle$, $\langle
0|\hat{\Lambda}^I\bar{H}_T^0=\langle 0|\hat{\Lambda}^IE_I$; recall that
$\hat{\mc{R}}^I=\hat{\pi}_T\hat{X}^I-\mr{i}\phi_I$. Similarly, the solution to the ground-state
problem takes the form: $\bar{H}_T^0|0\rangle=E_0|0\rangle$, and $\langle
0|\hat{\lambda}_0\bar{H}_T^0 = \langle 0|\hat{\lambda}_0E_0$. In addition, we define the projector: 
\begin{equation}
\hat{\mc{P}}_{IJ} = \begin{cases}
\tilde{r}_{JI}e^{\hat{T}}\hat{\mc{R}}^I|0\rangle\langle
0|\hat{\Lambda}^Je^{-\hat{T}} & \mr{if}~I>0,~J>0\\
\tilde{r}_{0I}e^{\hat{T}}\hat{\mc{R}}^I|0\rangle\langle 0|\hat{\lambda}_0 e^{-\hat{T}}
&\mr{if}~I>0,~J=0\\
\tilde{r}_{J0}e^{\hat{T}}|0\rangle\langle 0|\hat{\Lambda}^Je^{-\hat{T}} & \mr{if}~I=0,~J>0~,
\end{cases}
\end{equation}
where 
\ben
\tilde{r}_{JI}=(\mc{N}_{\mr{l},J}\mc{N}_{\mr{r},I})^{-1}~,
\een 
and, as noted before, $\tilde{r}_{II} = 1$. The projector can be used to study coherences and populations:
coherences are computed as $\langle \hat{\mc{P}}_{IJ}\rangle_{\Psi}(t)$ ($I\neq J$), and populations as
$\langle \hat{\mc{P}}_{II}\rangle_{\Psi}(t)$, where the projector is also inserted on the right hand side of
Eq. (\ref{gen_form}) for respective computation.

Let us consider three types of matrix elements, which we denote
$\tilde{\mc{B}}_{I0}$ ($I>0$), $\tilde{\mc{B}}_{0I}$ ($I>0$), and $\tilde{\mc{B}}_{IJ}$ ($I,J>0$).
The first transition element reads
\ben
\tilde{\mc{B}}_{I0} = \langle\hat{\Lambda}^I\bar{B}_T\rangle~.
\een
The computation of $\tilde{\mc{B}}_{I0}$ is more direct than the
others, it only involves $\hat{\Lambda}^I$. The counterpart of this element is
\ben\label{bt0I}
\tilde{\mc{B}}_{0I}=\langle \hat{\lambda}_0[\bar{B}_T,\hat{\pi}_T\hat{X}^I]\rangle
-\sum_{J>0}\frac{F^{IJ}}{\Omega_I+\Omega_J}\langle\hat{\Lambda}^J\bar{B}_T\rangle~,
\een
(note that $\mc{B}_{00}=\langle\hat{\lambda}_0\bar{B}_T\rangle$).
As in single-reference theory \cite{koch1990coupled,pedersen1997coupled}, transition
elements ($\mc{B}_{IJ}$, $I\neq J$) are not Hermitian conjugates of one another; and this is due to missing normalization
factors. Using the exact operators we find that
$\langle\hat{\lambda}_0[\bar{H}_T^0,\nabla\hat{\pi}_T\cdot\mb{X}^I\hat{X}^J]\rangle = 0$, and
\ben
\langle\hat{\lambda}^0\big[[\bar{H}_T^0,\hat{\pi}_T\hat{X}^I],\hat{\pi}_T\hat{X}^J\big]\rangle
=
-(\Omega_J+\Omega_I)\langle\hat{\lambda}_0(\hat{\pi}_T\hat{X}^I-\mr{i}\phi_I)(\hat{\pi}_T\hat{X}^J-\mr{i}\phi_{J})\rangle
+\Omega_I\langle\hat{\lambda}_0[\hat{\pi}_T\hat{X}^I,\hat{\pi}_T\hat{X}^J]\rangle~.
\een
Now we employ resolution of the identity, so
\ben\label{identi}
\sum_{I>0}\hat{\mc{P}}_{II}=
1-e^{\hat{T}}|0\rangle\langle 0|\hat{\lambda}_0e^{-\hat{T}}~.
\een
Under these expressions we have that
\ben
\sum_{J>0}\frac{F^{IJ}}{\Omega_I+\Omega_J}=
\langle\hat{\lambda}_0(\hat{\pi}_T\hat{X}^I-\mr{i}\phi_I)\bar{B}_T\rangle~.
\een
This result is constrained to the solution to all excitation orders, 
and it implies that $\tilde{\mc{B}}_{0I}= \langle
\hat{\lambda}_0\bar{B}_T(\hat{\pi}_T\hat{X}^I-\mr{i}\phi_I)\rangle$ in that limit.

For excited-state/excited-state matrix elements we examine:
\ben\label{btij}
\tilde{\mc{B}}_{IJ}=
\langle \hat{\Lambda}^I[\bar{B}_T,\hat{\pi}_T\hat{X}^J]\rangle
+\delta_{IJ}\langle\hat{\lambda}_0\bar{B}_T\rangle
+\sum_{K>0}\frac{\mc{G}_{IJK}}{\Omega_I-\Omega_J-\Omega_K}\langle\hat{\Lambda}^K\bar{B}_T\rangle~.
\een
This formula is identified from the previous subsection as well, Section \ref{unperturbed_TD_states}.
Assuming exact solution, several terms cancel in the calculation of $\mc{G}_{IJK}$. For
example, the term $\langle\hat{\Lambda}^I[\bar{H}_T^0,\nabla \hat{\pi}_T\cdot \mb{X}^J\hat{X}^K]\rangle$
cancels itself out with $\Omega_I\langle\hat{\Lambda}^I\nabla\hat{\pi}_T\cdot\mb{X}^J\hat{X}^K\rangle$. 
Expanding the first commutator on the right hand side of Eq. (\ref{gkij}), we find that
$\mc{G}_{IJK}=(\Omega_I-\Omega_J-\Omega_K)\langle\hat{\Lambda}^I\hat{\mc{R}}^J
\hat{\mc{R}}^K\rangle$.
Finally, inserting this result into Eq. (\ref{btij}) and using Eq. (\ref{identi}) give
$\tilde{\mc{B}}_{IJ}= \langle\hat{\Lambda}^I\bar{B}_T(\hat{\pi}_T\hat{X}^J-\mr{i}\phi_J)\rangle$. 

Having determined a relation between uncorrected transition amplitudes and CC wavefunctions, the formal connection
between CC and standard matrix elements is then expressed as:
\ben\label{BIJ}
\langle\Psi_I|\hat{B}|\Psi_J\rangle= \tilde{r}_{IJ} \tilde{\mc{B}}_{IJ}~.
\een
The asymmetric CC matrix representation of a given observable requires multiplication by
another asymmetric matrix in order to give the standard symmetric matrix. 
Eq. (\ref{BIJ}) also holds for the single-reference CC theory, where $\hat{\pi}_T=1$. In Sections \ref{numsim}
and \ref{discussion}, we discuss Eq. (\ref{BIJ}) step in a numerical context.  

\section{Numerical Simulation Details}\label{numsim}

\begin{figure}[htb!]
\centering
\includegraphics[scale=0.8]{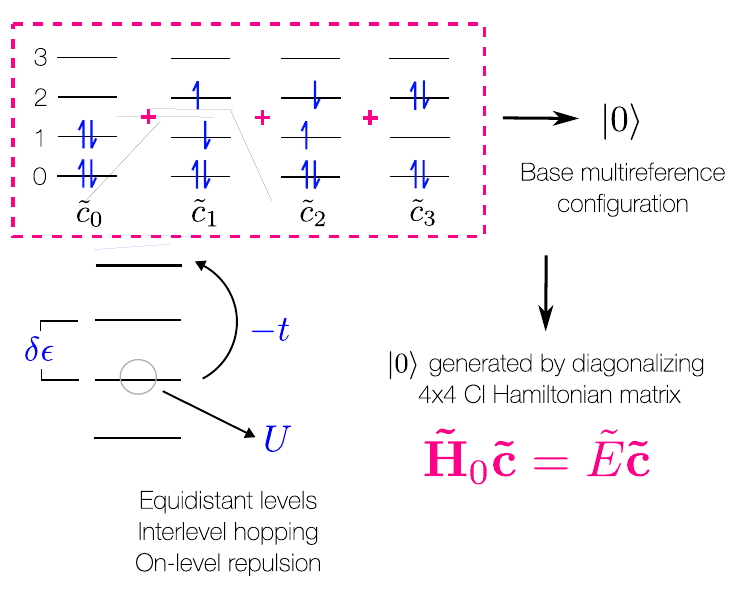}
\caption{Sketch of model: Four-level/four-electron system with on-level repulsion and inter-level
hopping. eTD-MRCC is performed to all particle excitation orders (up to quadruple
excitations).}\label{hub_model}
\end{figure}

We apply the approach developed in this work to a four-electron/four-level system, where the MR
state $|0\rangle$ is generated by diagonalizing the Hamiltonian in CI form over a (2,2)
configurational space. We introduce the Hubbard-like Hamiltonian operator of the system as:
\ben
\hat{H}_0 = \sum_{i\sigma} \epsilon_{i\sigma}\hat{a}^{\dagger}_{i\sigma}\hat{a}_{i\sigma}
-t\sum_{\mu\in S}(\tilde{\tau}_{\mu}+\tilde{\tau}^{\dagger}_{\mu})+U\sum_i
\hat{n}_{i\up}\hat{n}_{i\dn}~,
\een
where $\epsilon_{i\sigma} = i\times \delta\epsilon$ ($i\in\{0,1,2,3\}$), $\sigma$ denotes
electron-spin channel, and $\sum_{\mu\in S}$ means summation over all the possible conventional
single electron excitations. The TD version of this Hamiltonian is
$\hat{H}(t)=\hat{H}_0-\hat{D}f(t)$, where the dipole operator is taken as 
\ben
\hat{D}=D_0\sum_{\mu\in
S}(\tilde{\tau}_{\mu}+\tilde{\tau}_{\mu}^{\dagger})~,
\een
and the scalar driving field is a Gaussian pulse described by $f(t)=f_0\exp[-(t-t_0)^2/2\sigma^2]$. 
The code used to generate the data (Full MR CI and eTD-MRCC) is available at GitLab \cite{gitlab}.

Fig. \ref{hub_model} shows the four levels and the (2,2) configurational space. To generate the
reference we diagonalize their corresponding CI Hamiltonian, $\tilde{\mb{H}}_0$; this gives a vector
$\tilde{\mb{c}}$ (corresponding to the lowest-energy state) that we use to construct
$|0\rangle=\sum_N \tilde{c}_N|\mr{config}.~N\rangle$. To preserve the simplicity of the model
Hamiltonian, we do not optimize the molecular orbitals of the system. This procedure generates
$|0\rangle$ as the linear superposition of two doubly-occupied configurations, and two
singly-excited configurations. Then, initial states of interest are propagated under the action of
the scalar field $f(t)$. The model constants are $\delta\epsilon=1~\mr{eV}$, $U=0.250~\mr{eV}$,
$t=0.150~\mr{eV}$, $D_0 = 0.25~\mr{a.u.}$, $f_0=0.10~\mr{a.u.}$, $\sigma=50~\mr{a.u.}$ (or
approximately $1~\mr{fs}$), and $t_0 = 100~\mr{a.u.}$; the initial time is simply $t=0$.  

Unitary reference results were generated by first solving the full MR CI problem, or exact
diagonalization based on the orthonormal set $\{\hat{\tau}_N|0\rangle\}_{N\ge 0}$, and then
propagating the initial state of interest using the the midpoint rule $|\Psi(t+\delta
t)\rangle=\exp[-\mr{i}\delta t\hat{H}(t+\delta t /2)]|\Psi(t)\rangle$. The eTD-MRCC propagations
were performed with the Runge-Kutta 2nd order method (RK2), 8k steps in the range
$[0,350~\mr{a.u.}]$ (same number of steps applied to reference propagation). The observable
$D(t)=\langle \Psi(t)|\hat{D}|\Psi(t)\rangle$ is computed accordingly for each step. 
The MR state $|0\rangle$ is characterized by an occupation of the
second level ($i=1$) of $\sim 1.94$, and $\sim 0.06$ for the third level ($i=2$), reflecting a
significant MR character of the system. The CC operators account for the effect of all the
transitions neglected after the generation of $|0\rangle$, and produce results consistent with
standard quantum mechanics.

The algorithm used is entirely numerical, where we employ the three-index tensor based on
$G_{\mu\nu\rho}=\langle\mr{MR}|\tilde{\tau}_{\mu}^{\dagger}\tilde{\tau}_{\rho}
\tilde{\tau}_{\nu}|\mr{MR}\rangle$, and the contracted tensor: $g_{MNL}=\sum_{\mu\nu\rho}
Q^{\dagger}_{M\mu}Q_{\rho L}Q_{\nu N}G_{\mu\nu\rho}$. All operators are represented over the
orthonormal basis generated after the QR factorization, that is $\{\hat{\tau}_N|0\rangle\}$.  For
example, the Hamiltonian is expressed as $H_{MN}^0 =
\langle\hat{\tau}^{\dagger}_M\hat{H}_0\hat{\tau}_N\rangle$, and using the tensor $g_{MNL}$ we write
the operator $\hat{T}$ in this representation, so $\exp(\hat{T})$ is computed numerically in matrix
form; the same applies to the TD cluster operators. The GS variational problem is solved using the
quasi-Newton method, where a threshold value of $10^{-16}~\mr{a.u.}$ is employed for the energy
variations. Singular values are shown in the Supplemental Material \cite{SM}; they sharply drop from
$\sim 0.6$ down to less than $10^{-16}$, so the irrelevant singular vectors are clearly identified
and discarded. On ther other hand, the metric operator truncates after the fourth order nested commutator.
The numerical matrix representation of this operator is diagonally dominant, with off-diagonal
contributions resulting from the non-commutativity of conventional excitation operators. 

In terms of mathematical workflow, to complete an eTD-MRCC propagation of an non-ground initial state,
in our approach one first computes the cluster operators $\hat{x}(t)$ and $\hat{\lambda}(t)$, Eqs.
(\ref{dt_xm}) and (\ref{dt_lamb_m}) (without `m' subscript). The next step is determination of
$\hat{x}_{\mr{r}}(t)$, $\hat{\lambda}_{\mr{l}}(t)$, and $\hat{\lambda}_{\mr{lr}}(t)$, Eqs.
(\ref{xr_eq}), (\ref{lambl_eq}), and (\ref{lamb_lr}), which are then used to calculate the TD
observable through Eq. (\ref{gen_form}). Initial conditions ($t=0$) are computed following Eqs.
(\ref{lambs_t}), (\ref{l_init}), (\ref{eq_for_D}), (\ref{dlr}), and (\ref{CI_DI}), these require
solution to the EOM-MRCC problem. Because the motion equation are local in time, all the cluster
operators can be determined together within the same discretized propagation step. For example, to
calculate the operators at $t+\delta t$, one just needs evaluation and estimation of quantities in
the $[t,t+\delta t]$ interval, as in a conventional Runge-Kutta problem. The numerical matrix
algebra (`double precision') was carried out with the `Eigen' C++ library \cite{eigenweb}, and the
configurations are represented using `Boost' dynamical bitstrings \cite{BoostLibrary}.

\section{Numerical Results and Discussion}\label{discussion}

\begin{figure}[htb!]
\centering
\includegraphics[scale=1.0]{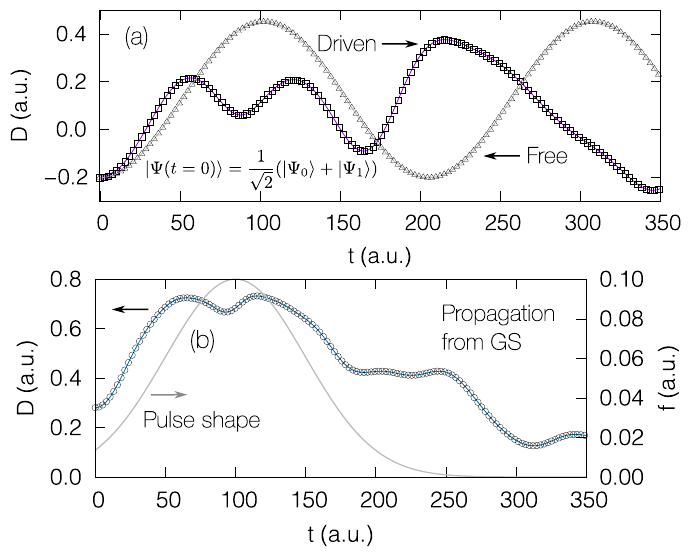}
\caption{Time dependency of observable $\hat{D}$. Solid lines: eTD-MRCC, symbols: numerically
exact, reference: unitary propagation. (a) Propagation from a superposition state: Open squares
represent propagation subject to the external field, $f(t)$, and open triangles correspond to free propagation (field
turned `off'). Matching lines correspond to eTD-MRCC, purple to driven propagation, and grey to free
propagation. (b) TD dipole for evolution from GS and shape of pulse used (grey). Open circles represent UP
results, and matching solid (black) line eTD-MRCC.}\label{td_dip}
\end{figure}

As mentioned, we perform full MR CI calculations (exact diagonalization over the basis
$\{\hat{\tau}_N|0\rangle\}$) that are used as reference to verify that our model produces numerically
equivalent GS and ES energies, up to the 15th decimal digit (Supplemental Material \cite{SM}). It is
expected, as MRCC derives from exact quantum mechanics. The same applies to EOM-MRCC, and our
generalized eigenvalue problem adds the metric operator, but the term $\hat{\pi}_T\hat{X}^I$ can be
treated like a regular excitation operator. In this eTD-MRCC work, however, the explicit inclusion
of this metric is convenient as the rest of the formalism needs the time dependency of this
operator. Numerical uncertainty at the fine level may be attributed to the eigensolver, SVD
decomposition, time propagation method, and exponential matrix calculator.

Having solved the EOM-MRCC problem, one then constructs the initial state and propagates it.  We
analyze the results for the propagation of the initial state
$|\Psi(0)\rangle=1/\sqrt{2}(|\Psi_0\rangle+|\Psi_1\rangle)$, where $|\Psi_1\rangle$ represents the
first singlet excited state of this four-level/four-fermion system, and compare against propagation
from the GS alone, $|\Psi(0)\rangle=|\Psi_0\rangle$. Fig. \ref{td_dip}(a) and \ref{td_dip}(b) show
the performance of eTD-MRCC with reference to the conventional unitary propagation (UP) for the
observable $D(t)=\langle\Psi(t)|\hat{D}|\Psi(t)\rangle$. In case of GS propagation, we note that Eqs.
(\ref{dt_xm}) and (\ref{dt_lamb_m}) produce the same results as the reference UP, Fig.
\ref{td_dip}(b). After application of the ultrafast pulse, $D(t)$ displays
a non-linear behavior where contributions from different frequencies can be noted. The dipole
then continues behaving in an non-stationary fashion, as expected for a coherent system. For propagation of
$|\Psi(0)\rangle=1/\sqrt{2}(|\Psi_0\rangle+|\Psi_1\rangle)$, Fig. \ref{td_dip}(a), we also observe
a relatively fast response to the pulse, but with more defined oscillatory `signals', which are
due to the superposition nature of the initial state (as opposed to starting exclusively from the GS).
The eTD-MRCC results reproduce the reference UP in an excellent manner.
The response of the system under this initial state is considerably strong in comparison with
the propagation in absence of pulse, Fig. \ref{td_dip}(a), where the dipole is a harmonic function
with a periodic time of about $411~\mr{a.u.}$.

\begin{figure}[htb!]
\centering
\includegraphics[scale=1.0]{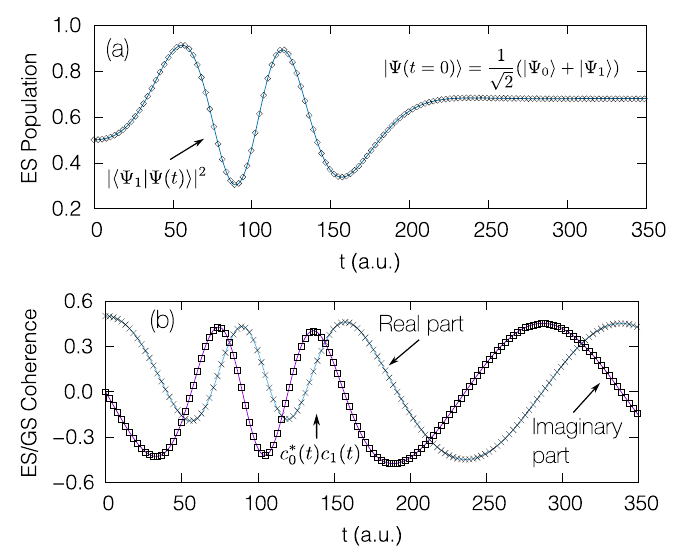}
\caption{Excited state TD probability, (a), and coherence, (b). Solid lines: eTD-MRCC, symbols:
numerically exact, reference, unitary propagation. In (a), open diamonds represent UP propagation results
for ES population, and matching solid line (light blue) corresponds to eTD-MRCC. (b) Real and imaginary
parts of coherence, open squares and matching (purple) solid line refer to imaginary component, and
`x' symbols with their matching (light blue) line correspond to UP and eTD-MRCC.}\label{ps_cohs}
\end{figure}

The developed eTD-MRCC formalism offers a relatively simple symmetrization matrix that upon
multiplication by the conventional CC transition element, it corrects it to match that of standard
quantum mechanics. For the  matrix representation of $\hat{D}$, the numerical simulation gives the
same value up to the 13th decimal digit (Supplemental Material \cite{SM}). In RK2 propagation with
medium-quality time steps, we note small asymmetric numerical errors in the matrix, but these are
mainly due to the RK2 solver errors, because reducing the time-step systematically lowers these
deviations. The eTD-MRCC approach predicts population and coherences. In Fig. \ref{ps_cohs} we
observe that eTD-MRCC is numerically producing ES population $|\langle\Psi_1|\Psi(t)\rangle|^2$ and
coherence $c_0^*(t)c_1(t)$ ($c_I(t)=\langle\Psi_I|\Psi(t)\rangle$) profiles in full agreement with the
reference UP. So eTD-MRCC is promising in that it could allow for the quantitative examination, from
a CC perspective, of eigenstate probabilities and interferences between states. 

\begin{figure}[htb!]
\centering
\includegraphics[scale=1.0]{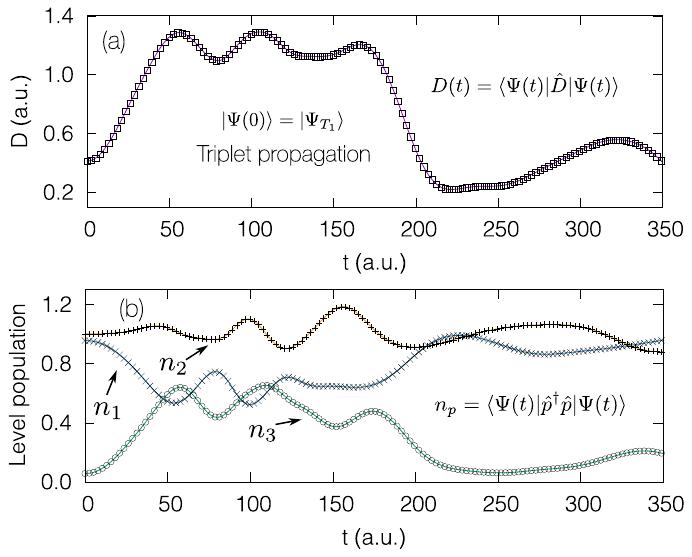}
\caption{Results for propagation from lowest triplet excited state. Solid lines: eTD-MRCC, symbols:
numerically exact, reference, unitary propagation. (a), Propagation from first triplet state, open
squares: UP, matching (purple) line: eTD-MRCC. (b), Level populations as a function of time; `x'
symbols, `+' symbols, and open circles represent populations of the second, third, and fourth
level, correspondingly, and matching lines (blue, orange, and green) to their eTD-MRCC results. 
Driving pulse is the same used before for singlet states.}\label{triplet}
\end{figure}

Fig \ref{triplet}(a) displays the results for the propagation of the lowest triplet state,
$|\Psi(0)\rangle=|\Psi_{T_1}\rangle$. This state is treated as an excited state that derives from
the EOM-MRCC solution, and is slightly
below the first excited singlet state.  We observe that
eTD-MRCC produces consistent results for this propagation.  The time-dependency of the level
populations of the second and their levels are presented in Fig. \ref{triplet}(b), where the fourth
level ($i=3$) gains population in response to the pulse, and the second level ($i=1$) loses
population. The third level displays relatively small oscillations, but it overall does not change
as much as the second and fourth level do.

So far, we have discussed `numerically exact' propagations, but for practical calculations cluster
operators are truncated. In such case, even with the resymmetrization factors the resulting
observable matrices remain strictly non-symmetrical, which is due to the fundamental differences
between the left- and right-handed expressions. One possibility is to either arithmetically or
geometrically average non-diagonal matrix elements in order to construct Hermitian TD matrices, or
use an advanced form of symmetrization. As a further approximation, motivated by the fact that
$\mc{N}_{\mr{r},I}\mc{N}_{\mr{l},I}=1$, it is possible to take all resymmetrization factors as
the unity. However, explicitly including $\tilde{r}_{IJ}$ might support a systematic computation
of matrix elements and TD observables. 

The contractions of conventional particle/hole operators (which produce $\{\hat{\tau}_N\}$) used in
this work facilitate the derivation of a formalism with analogies to the single-reference scenario.
Furthermore, the present theoretical framework is compatible with ic-MRCC theories in which the
reference is optimized together with the GS $\hat{T}$ operator, giving an appropriate form for
$|0\rangle$. Besides added numerical stability, such optimization is advantageous because it reduces
the size of the transformations needed to generate cluster operators. In other words, contracted
excitation operators are dependent upon choice of multireference (a subject to be investigated in
the future from an efficiency perspective for TD propagations). Of future interest is also the
extension of our formalism to account for TD relaxation of orbitals. As in GS ic-MRCC theory, TD
relaxation could be numerically convenient for the determination of the associated TD cluster
amplitudes, and would require additional mathematical elements in the formalism and its algorithmic
implementation.

Regarding density matrices, the procedure described in this work applies to sample a pure density
matrix. For mixed-state density matrices, one has to consider the collection of possible TD cluster
operators. For example, propagation of two different pure excited states demands for two different
sets of $\hat{x}_{\mr{r}},~\hat{\lambda}_{\mr{l}},~\hat{\lambda}_{\mr{lr}}$ operators. So a density
matrix framework of mixed states requires analysis of averages over an ensemble of extended TD
cluster operators, and the equation of motions derived in this work. Appendix \ref{dm_tdeq} defines
auxiliary density matrices, and their motion equations, which might assist in the modeling of the
coupling between quantum systems and the environment, and the study of dissipative effects.

\section{Conclusions}
For propagation of general initial states, this work presented an internally contracted time-dependent multireference
coupled-cluster formalism based on the infinitesimal analysis of modified cluster operators, which
are non-commutative in this context. Main developments reported in this work are summarized as
follows: i) We used our previously proposed `second response' theory to derive an excited-state
generalized eigenvalue problem that is similar to the conventional EOM-CC approach, but it involves
a metric operator, which is needed for the rest of the formalism. ii) CC motion equations were found
for propagations of general initial states such as pure ground or excited states, and linear
combinations of them. As mentioned before, the resulting formalism is size extensive.  iii)
Based on linear-response theory and analysis of coherent free (unperturbed) propagations,
formulas were derived for the matrix representations of general observables, including a
resymmetrization matrix that in principle turn them into exact quantities. Finally, motivated by
the numerical model, these three advances suggesst that, in the exact limit, there is
equivalence between eTD-MRCC and unitary quantum mechanics.

\appendix
\section{Derivation of Equation (3)}\label{der_rule}
For a given TD cluster operator $\hat{u}(t)$, we proceed to show that:
\ben
\frac{\partial\hat{\pi}_{u}}{\partial t}\hat{\tau}_N-\frac{\partial\hat{\pi}_u}{\partial
u_N}\frac{\partial \hat{u}}{\partial t} =
\Big[\hat{\pi}_u(t)\hat{\tau}_N,\hat{\pi}_u(t)\frac{\partial \hat{u}}{\partial t}\Big]~.
\een
Let us consider the function
\ben
F(\hat{u}(t))= \exp(-\hat{u})\frac{\partial}{\partial
u_N}\exp(\hat{u})=\hat{\pi}_u\hat{\tau}_N~.
\een
Using the chain rule yields:
\ben\label{chain}
\frac{\partial F}{\partial t} = \sum_M \frac{\partial F}{\partial u_M} \frac{\partial
u_M}{\partial t}~,
\een
and using the identity $\partial_{\alpha}(e^{-\hat{u}}\partial_{\beta}e^{\hat{u}})=
[\hat{\pi}_u\partial_{\beta}\hat{u},\hat{\pi}_u\partial_{\alpha}\hat{u}]+\partial_{\beta}(\hat{\pi}_u\partial_{\alpha}\hat{u})$,
we find:
\ben
\frac{\partial}{\partial u_M}\Big(e^{-u}\frac{\partial}{\partial u_N}e^{+u}\Big)= 
[\hat{\pi}_u\hat{\tau}_N,\hat{\pi}_u\hat{\tau}_M]+\frac{\partial}{\partial
u_N}\Big(e^{-u}\frac{\partial}{\partial u_M}e^{+u}\Big)~.
\een
Alternatively,
\ben
\frac{\partial\hat{\pi}_u}{\partial u_M}\hat{\tau}_N= 
[\hat{\pi}_u\hat{\tau}_N,\hat{\pi}_u\hat{\tau}_M]+\frac{\partial\hat{\pi}_u}{\partial
u_N}\hat{\tau}_M~.
\een
Through Eq. (\ref{chain}) we obtain
\ben
\frac{\partial\hat{\pi}_u}{\partial t}\hat{\tau}_N =
\Big[\hat{\pi}_u(t)\hat{\tau}_N,\hat{\pi}_u(t)\frac{\partial\hat{u}}{\partial t}\Big]
+\frac{\partial\hat{\pi}_u}{\partial u_N}\frac{\partial \hat{u}}{\partial t}~.
\een
This completes the proof.

\section{$\lambda_{\mr{r}}$ TD Equation}\label{appdx_lambr}
After differentiating Eq. (\ref{dt_lamb_m}), the operator $\hat{\lambda}_{\mr{r}}$ is described by the TD equation:
\ben\label{r_eq}
\begin{split}
-\mr{i}\langle\partial_t\hat{\lambda}_{\mr{r}}\hat{\pi}_x(t)\hat{\tau}_N\rangle=&
\langle\hat{\lambda}_{\mr{r}}(t)[\hat{H}_x(t),\hat{\pi}_x(t)\hat{\tau}_N]\rangle+
\langle\hat{\lambda}(t)\big[[\hat{H}_x(t),\hat{\pi}_x(t)\hat{x}_{\mr{r}}(t)],\hat{\pi}_x(t)\hat{\tau}_N\big]\rangle\\
&+\langle\hat{\lambda}(t)[\hat{H}_x(t),\hat{\pi}_r(t)\hat{\tau}_N]\rangle
+\Gamma_{\mr{r},N}(t)~,
\end{split}
\een
where
\ben
\begin{split}
-\mr{i}&\Gamma_{\mr{r},N}(t) = 
\langle \hat{\lambda}_{\mr{r}}(t)[\hat{\pi}_x(t)\hat{\tau}_N,\hat{\pi}_x(t)\partial_t\hat{x}]\rangle+
\langle \hat{\lambda}(t)[\hat{\pi}_{\mr{r}}'(t)\hat{\tau}_N,\hat{\pi}_x(t)\partial_t\hat{x}]\rangle\\
&+\langle \hat{\lambda}(t)[\hat{\pi}_x(t)\hat{\tau}_N,\hat{\pi}_{\mr{r}}'(t)\partial_t\hat{x}]\rangle+
\langle \hat{\lambda}(t)[\hat{\pi}_x(t)\hat{\tau}_N,\hat{\pi}_x(t)\partial_t\hat{x}_{\mr{r}}]\rangle
+\langle \partial_t\hat{\lambda}\hat{\pi}_{\mr{r}}'(t)\hat{\tau}_N\rangle~.
\end{split}
\een

\section{Computation of $\pi_{\mr{r}}$}\label{pir_section}
This object can be computed recursively, which is noticed by starting from the product rule
of differentiation:
\ben
\mr{d}(\mr{ad}_XZ) = \mr{ad}_{\mr{d}X}Z +\mr{ad}_X \mr{d}Z~,
\een
where $\mr{d}$ is a differential operator.
Setting $Z = \mr{ad}^{n-1}_X Y$, where $Y$ is constant, gives
\ben
\mr{d}(\mr{ad}_X^n Y) = \mr{ad}_{\mr{d}X}\mr{ad}_X^{n-1}Y +\mr{ad}_X \mr{d}(\mr{ad}^{n-1}_X)Y~,
\een
resulting in a convenient formula for objects such as $\hat{\pi}_{\mr{r}}'(t)$. 

\section{Ensemble Averaging}\label{dm_tdeq}
Consider an ensemble, $\Gamma$, of systems that start at different initial states. For each initial state,
there is a group of three TD cluster operators, we denote the whole set of operators as
$\{\hat{x}_{\mr{r}}^k,\hat{\lambda}_{\mr{l}}^k, \hat{\lambda}_{\mr{lr}}^k\}$, where $k$ labels a
particular type of ensemble member. The average of an observable is thus:
\ben
\langle \hat{B}\rangle_{\Gamma}(t) = \sum_k \big\{w_k
\langle\hat{\lambda}_{\mr{l}}^k(t)\hat{b}_x(t)\hat{\pi}_x(t)\hat{x}_{\mr{r}}^k(t)\rangle + 
w_k \langle\hat{\lambda}_{\mr{lr}}^k(t)\hat{B}_x(t)\rangle\}~,
\een
where $\{w_k\}$ are the weights, and $\hat{b}_x(t)$ acts over the cluster operator on its right 
as $\hat{b}_x(t)\hat{z}|0\rangle = [\hat{B}_x(t),\hat{z}]|0\rangle$.
Motivated by the above equation, we define the auxiliary density operators:
\ben
\begin{split}
\hat{\gamma}_1(t) &= \sum_k w_k \hat{x}_{\mr{r}}^k(t)|0\rangle\langle 0|\hat{\lambda}_{\mr{l}}^k(t)~,\\
\hat{\gamma}_2(t) &= \sum_k w_k \hat{\lambda}_{\mr{lr}}^k(t)~.
\end{split}
\een
Using these objects we have that:
\ben
\langle \hat{B}\rangle_{\Gamma}(t) = \mr{tr}\{\hat{b}_x(t)\hat{\pi}_x(t)\hat{\gamma}_1(t)\} + \langle
\hat{\gamma}_2(t)\hat{B}_x(t)\rangle~.
\een
The operator $\hat{\gamma}_2(t)$ follows the same motion equation as $\hat{\lambda}_{\mr{lr}}(t)$.
Through Eqs. (\ref{xr_eq}), (\ref{lambl_eq}), and Appendix \ref{der_rule} (for variations of
$g_{\mr{R}}$ instead of $t$), we find for $\hat{\gamma}_1(t)$ that:
\ben\label{gam_evol}
\mr{i}\hat{\pi}_x(t)\frac{\mr{d}\hat{\gamma}_1}{\mr{d}t}\hat{\pi}_x(t)
  = [\hat{h}_x(t),\hat{\pi}_x(t)\hat{\gamma}_1(t)]\hat{\pi}_x(t)+\hat{\varphi}(t)~,
\een
where $\hat{h}_x(t)\hat{z}|0\rangle=[\hat{H}_x(t),\hat{z}]|0\rangle$, and 
\ben\label{hatphi}
\mr{i} \hat{\varphi}(t)
=\nabla\hat{\pi}_x\cdot \dot{\mb{X}}(t)\hat{\gamma}_1(t)\hat{\pi}_x(t)
+\hat{y}_x(t)\hat{\gamma}_1(t)\hat{\pi}_x(t)
-\hat{\pi}_x(t)\hat{\gamma}_1(t)\hat{y}_x(t)~,
\een
where
$\hat{y}_x(t)\hat{z}|0\rangle=[\hat{\pi}_x(t)\partial_t\hat{x},\hat{\pi}_x(t)\hat{z}]|0\rangle$.
The object $\nabla\hat{\pi}_x\cdot \dot{\mb{X}}(t)$ follows the same relation as Eq.
(\ref{nabpi_t}), with $\hat{T}$ replaced by $\hat{x}(t)$, and $\mb{X}^I$ by $\dot{\mb{X}}(t)$.  For
numerical purposes, Eqs. (\ref{gam_evol}) and (\ref{hatphi}) have to be converted to matrix
representation, that is $\langle \hat{\tau}_M^{\dagger}\hat{\gamma}_1(t)\hat{\tau}_N\rangle$ and
$\langle \hat{\tau}_M^{\dagger}\hat{\varphi}(t)\hat{\tau}_N\rangle$.

\begin{acknowledgments}
This material is based upon work supported by the U.S. Department of Energy, Office of Science,
Office of Basic Energy Sciences Early Career Research Program under Award Number DE-SC-0025662.
\end{acknowledgments}

\bibliography{doe_refs}

\end{document}